\documentclass[prb,aps,showpacs,twocolumn,preprintnumbers,amsmath,amssymb,superscriptaddress]{revtex4-1}
\usepackage{amsmath,amssymb,amsfonts}
\usepackage{graphicx}
\usepackage[colorlinks=True,linkcolor=red,citecolor=blue,urlcolor=blue]{hyperref}
\usepackage{xcolor}
\usepackage{chngcntr}
\usepackage{braket}
\usepackage{nicefrac}
\usepackage{bm}
\usepackage{multirow}
\usepackage{blkarray}
\usepackage{comment}
\usepackage{cleveref}
\usepackage{pifont}
\usepackage{colortbl}
\usepackage[normalem]{ulem}

\crefname{equation}{Eq}{Eqs} 
\crefrangelabelformat{equation}{(#3#1#4--#5#2#6)}

\def\ket#1{|#1\rangle }
\def\bra#1{\langle #1 |}

\newcommand{\FJ}[1]{{\color{black} #1}}

\newcommand{\DM}[1]{{\color{black} #1}}

\newcommand{\edits}[1]{{\color{black}  #1}}

\begin{document}
\bibliographystyle{apsrev4-1}
\title{Optical probes of two-component pairing states in transition metal dichalcogenides}

\author{Miguel-\'Angel S\'anchez-Mart\'inez}
\affiliation{H. H. Wills Physics Laboratory, Tyndall Avenue, Bristol, BS8 1TL, United Kingdom}
\author{Daniel Muñoz-Segovia}
\affiliation{Donostia International Physics Center, 20018 Donostia-San Sebastian, Spain}
\affiliation{Department of Physics, Columbia University, New York, NY 10027, USA}
\author{Fernando de Juan}
\affiliation{Donostia International Physics Center, 20018 Donostia-San Sebastian, Spain}
\affiliation{IKERBASQUE, Basque Foundation for Science, Plaza Euskadi 5, 48009 Bilbao, Spain}

\date{\today}
\begin{abstract}

\DM{Signatures of unconventional superconductivity have been recently observed in certain transition metal dichalcogenides (TMDs), including 4H$_b$-TaS$_2$ and monolayer 2H-NbSe$_2$}. While the pairing channel remains unknown, it has been argued that spin fluctuations \FJ{can stabilize pairing in the two-component $E'$ channel,} \DM{a $p$-wave spin-triplet state} \FJ{which could be consistent with some of the reported signatures.} 
Exploiting the particular multi-orbital character of the Fermi surface and the presence of Ising spin-orbit coupling, which enable finite optical conductivity in the clean limit, in this work we predict clear-cut optical signatures \DM{to detect and distinguish} the chiral and nematic ground states of the $E'$ pairing. 
\FJ{We quantify how nematic $E'$ states produce a diagonal anisotropy $\sigma_{xx}\!\neq\!\sigma_{yy}$ due to the broken threefold symmetry ($C_3$), while chiral $E'$ states yield a finite optical Hall conductivity $\sigma_{xy}^H$ due to broken time-reversal symmetry, and find both signals could be detected in current experiments. \DM{ For instance, for realistic gaps in the meV range, we predict a relative anisotropy $\Delta\sigma/\sigma\sim10^{-5}$ in the nematic states, and a polar Kerr rotation of $\theta_K\!\sim\!10^{-5}$ rad} in the chiral states. These symmetry fingerprints provide a practical route to distinguish nematic and chiral superconducting order in TMD superconductors.}

\end{abstract}
\maketitle

\section{Introduction}
Metallic transition metal dichalcogenides (TMDs) MX$_2$ (M=Nb,Ta, X=S, Se)~\cite{Wilson75}, believed to be standard BCS superconductors in their bulk 2H polytype, have recently shown signatures of unconventional superconductivity~\cite{Norman11} when made into thin films and heterostructures, calling for a better understanding of their pairing symmetries and mechanisms. For example, NbSe$_2$ thin films display spontaneous breaking of threefold symmetry in the superconducting state under in-plane magnetic fields 
~\cite{Hamill21,Cho22}, as well as superconducting collective modes \cite{Wan22} and anomalous resilience to magnetic fields \cite{Kuzmanovic22}. \FJ{Evidence for topological edge states, which originate from unconventional pairing}, has also been reported in monolayer CrBr$_3$ on bulk NbSe$_2$ \cite{Kezilebieke20}. 4H$_b$-TaS$_2$~\cite{DiSalvo73}, which can be seen as a heterostructure of H and T polytypes, \FJ{shows anomalous muon spin rotation at the critical temperature $T_c$ which has been claimed to originate from time-reversal symmetry (TRS) breaking\cite{Ribak20}. Other experiments reported} spontaneous vortices in the superconducting state~\cite{Persky22}, superconducting edge modes \cite{Nayak21} and anomalous Little-Parks oscillations and upper critical field anisotropy \cite{Almoalem24,Silber24}. A chiral superconducting state has also been reported in 2H-TaS$_2$ intercalated with chiral molecules~\cite{Wan23}. \FJ{Since a two-component order parameter may condense into nematic or chiral superconducting states, thus breaking threefold and time-reversal symmetries, respectively, these experiments represent tantalizing indirect evidence of two-component pairing. Demonstrating its existence as a ground state stable phase, however, remains a major challenge, because so far the evidence of broken TRS is only indirect and broken threefold symmetry only occurs in the presence of an applied field.}


\FJ{The potential candidate channels for unconventional pairing in these TMD systems are} strongly constrained by the presence of Ising spin-orbit coupling (SOC)~\cite{Xi16,Sohn18,Barrera18,Wickramaratne20}, 
induced by a crystal structure which breaks inversion symmetry but preserves a horizontal mirror $\sigma_h$. The pairing problem for H-TMDs has been studied at length~\cite{He18,Mockli18,Shaffer20,Margalit21,Dentelski21,Horhold23,Liu24Interlayer}, in particular in the presence of two-component order parameters \cite{Hsu17,Chen19,Lane22}, but without identifying a microscopic interaction that makes these channels attractive. 
Previous works had however identified that monolayer H-TMDs are near a magnetic instability towards an antiferromagnetic state with in-plane polarization~\cite{Zheng18,Wickramaratne20,Das21,Costa22}, which leads to strong transverse paramagnons that can mediate unconventional superconductivity. A more recent calculation reveals that such fluctuations favor the two-component $E'$ triplet state in the presence of Ising SOC \cite{Roy24,Siegl24}, although it remains to be established whether this solution is robust in the presence of additional electron-phonon coupling \cite{Das23,Roy25}. \FJ{This $E'$ channel, which pairs electrons of opposite spins, is a $p$-wave triplet unaffected by Ising SOC and can condense in nematic $(p_x,p_y)$ or chiral $p_x+i p_y$ ground states. It is thus a strong candidate to explain the phenomenology observed in TMD systems.}

\FJ{Since the evidence of TRS and threefold symmetry breaking in these TMDs is so far indirect, an experimental probe that is directly sensitive to both symmetries in the superconducting state would be desirable to establish the existence of the $E'$ ground state. In this work, we choose the linear optical conductivity as such a probe. Indeed, it is well known that conductivity anisotropy $\sigma_{xx}\!\neq\!\sigma_{yy}$, which leads to observable features like linear dichroism and birrefringence, becomes enabled when perpendicular rotation axes $C_n$ with $n>2$ are absent~\cite{Rodger97,Newnham04,Mirri16,Xu22}, like in nematic states. In addition, Onsager relations dictate that the antisymmetric (or Hall) part of the conductivity, which leads to magnetic circular dichroism and polar Kerr effect~\cite{argyres1955theory,Kapitulnik2006,Kapitulnik2009,schemm14,Taylor12,Wang14,Wang17}, is only allowed when time-reversal and vertical mirror symmetries are broken, like in the chiral superconducting state. Optical conductivity is therefore ideally suited to probe the $E'$ state.}


\begin{figure}
    \centering
    \includegraphics[width=\linewidth]{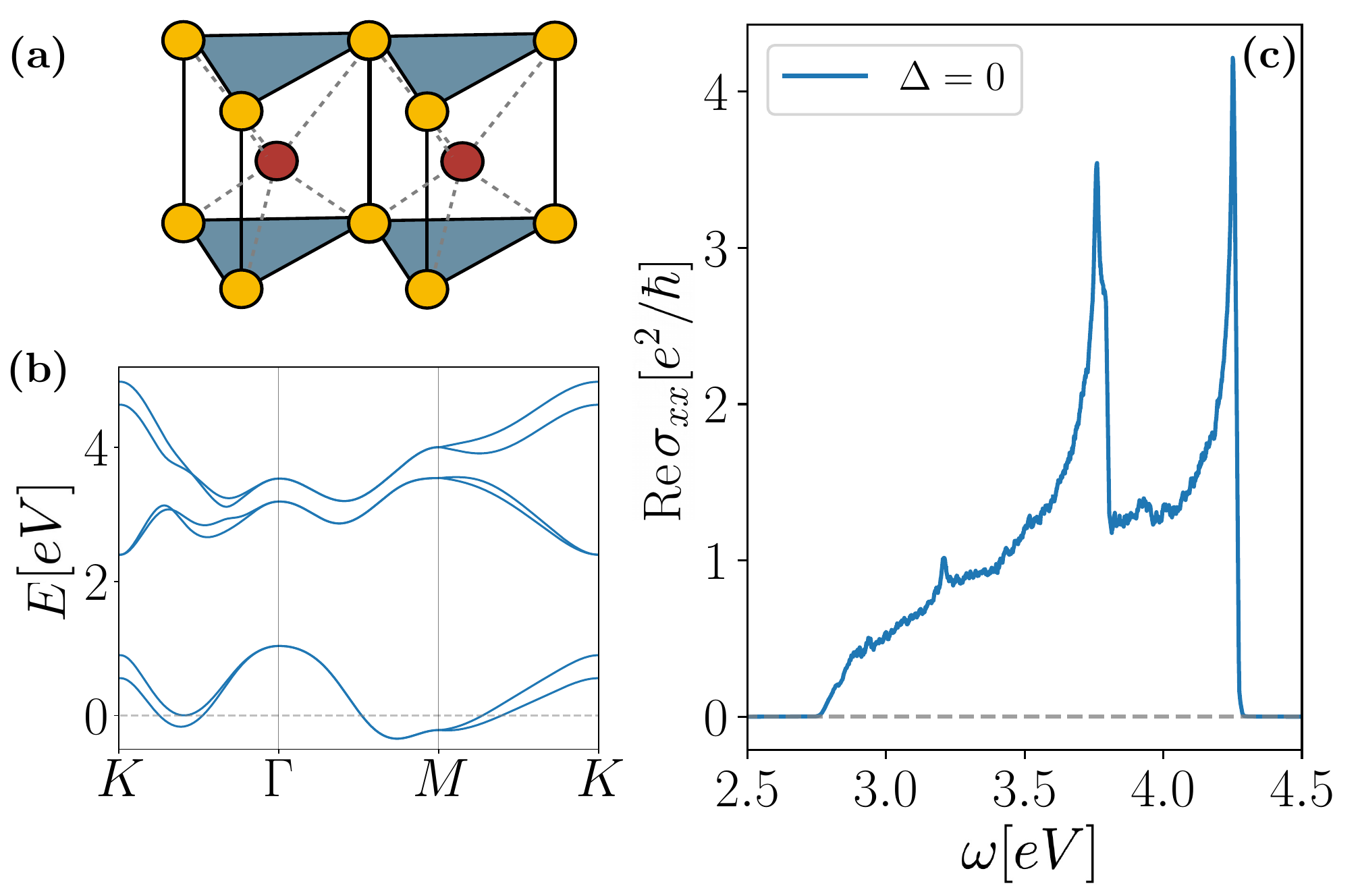}
    \caption{H-polytype of TMDs. (a) Crystalline structure of a single H-layer, corresponding to an ABA stacking of hexagonal lattices. (b) Band structure of the normal state of the H layer of TaS$_2$. (c) Optical conductivity in the normal state of TaS$_2$ shown in the range where it is nonzero.}
    \label{fig:Hlayernormalstate}
\end{figure}


\FJ{
Historically, the optical conductivity of superconductors has been understood in the Mattis--Bardeen, single-band dirty limit, which is essentially disorder-driven and leads to vanishing conductivity in the absence of scattering. But more recently it has been realized that clean multiband superconductors have finite optical conductivity~\cite{Ahn21}, which is sensitive to the pairing channel and is connected to the superfluid weight~\cite{Ahn21a,Chen21}. Previous work has considered the non-linear optical response of non-centrosymmetric superconductors \cite{Xu19,Tanaka23}, and recently both linear and non-linear conductivities were computed for the H-TMD three band model \cite{Raj24} with the single-component pairing states proposed in Ref.~[\onlinecite{Margalit21}]. The $E'$ state~\cite{Roy24,Siegl24} was not considered.  }

In this work, we consider the optical conductivity of the $E'$ superconducting state in both its chiral and nematic ground states, demonstrating that the optical Hall conductivity $\sigma_{xy}^H$ and the conductivity anisotropy $\sigma_{xx}-\sigma_{yy}$, which are zero for all other superconducting states, serve as unique probes for the chiral and nematic $E'$ states, respectively. 

\section{Model and pairing channels}
The H polytype of monolayer TMDs MX$_2$ (M=Nb,Ta, X=S, Se) crystallizes in a trigonal prismatic crystal structure (Fig~\ref{fig:Hlayernormalstate} (a)) with $D_{3h}$ point-group symmetry. The Fermi surface derives from ${d_{z^2},d_{x^2-y^2},d_{xy}}$-like orbitals centered at the transition metal positions\cite{Mockli18}. We consider a spinful three-band tight-binding model with up to third-nearest neighbours hoppings\cite{Liu13} in the orbital basis $\{d_{z^2},d_{x^2-y^2},d_{xy}\}$ which reads
\begin{align}
h_{\boldsymbol{k}}
 =\left[\begin{array}{cc}
h^{\mathrm{TNN}}_{\boldsymbol{k}}+\frac{\lambda}{2} L_z & 0 \\
0 & h^{\mathrm{TNN}}_{\boldsymbol{k}}-\frac{\lambda}{2} L_z
\end{array}\right],
\end{align}
where $h^{\mathrm{TNN}}_{\boldsymbol{k}}$ is the tight-binding Hamiltonian with up to third nearest-neighbours hoppings, $\lambda$ is the SOC strength, and $L_z$ is the matrix of the z-component of the angular momentum in the orbital basis $\{d_{z^2},d_{x^2-y^2},d_{xy}\}$. See Ref.~[\onlinecite{Liu13}] for parameter details and Fig.~\ref{fig:Hlayernormalstate}(b) for the resulting band structure.

Superconductivity is described with the Bogoliubov-de Gennes (BdG) Hamiltonian defined in terms of Nambu spinors $\Psi = (\psi,\psi^{\dagger})^T$ as 
\begin{align}
\mathcal{H} = \frac{1}{2} \Psi^{\dagger}
H_{\rm BdG} \Psi= \frac{1}{2} \Psi^{\dagger}_{\boldsymbol{k}}
\left(\begin{array}{cc} h_{\boldsymbol{k}} & \Delta_{\boldsymbol{k}} \\ \Delta_{\boldsymbol{k}}^\dagger &  - h_{-\boldsymbol{k}}^* 
\end{array}\right) \Psi_{\boldsymbol{k}}
\label{eqn:bdg_hamiltonian}
\end{align}
with $\Delta^\dagger_{\boldsymbol{k}} = - \Delta_{-\boldsymbol{k}}^*$ the superconducting gap. For simplicity we consider only momentum-independent pairing states where $\Delta_{\boldsymbol{k}} = \Delta$ is a 6x6 matrix in spin and orbital space. 

The different pairing channels are classified as irreducible representations (irreps) of the $D_{3h} = M_z \otimes C_{3v}$ point group, generated by mirror planes perpendicular to the $x$-axis $M_x$ and $z$-axis $M_z$, and threefold rotations $C_3$ around the $z$-axis. This group contains the $M_z$-even irreps $A_1'$, $A_2'$, $E'$ and $M_z$-odd irreps $A_1''$, $A_2''$, $E''$. Since the orbitals that contribute to the bands at the Fermi level are all $M_z$-even and we only consider intralayer pairing, odd $M_z$ parity can only come from the spin part of the pairing. Singlet pairing $\Delta = i \sigma_y$ and out-of-plane triplet $\Delta = i\sigma_y \sigma_z = \sigma_x$ are $M_z$-even pairings (which correspond to opposite spin pairing), while in-plane triplets $\Delta = i\sigma_y (\sigma_x,\sigma_y)$ are $M_z$-odd pairings (which correspond to equal-spin pairing). Since $D_{3h}$ does not contain the inversion operator, singlet and triplet channels can generally mix due to SOC, and channels are only classified by the global (spin and orbital) symmetry of the pairing.

Previous works have discussed different unconventional equal-spin pairing channels driven by the breaking of $M_z$. An equal-spin pairing chiral $p+ip$ superconducting state was predicted in the presence of large Rashba spin-orbit coupling~\cite{Shaffer20}, while Ref.~[\onlinecite{Margalit21}] proposed an equal-spin pairing one-dimensional irrep $A_2''$. Extending beyond these works, Ref.~[\onlinecite{Liu24Interlayer}] proposed an interlayer spin-polarized pairing state with odd parity. Given the large Ising SOC, which suppresses equal-spin pairing, $M_z$-even states are however more likely to be realized unless the scale of $M_z$ breaking is as large as Ising SOC. Although the $A_1'$ channel, which corresponds to the standard $s$-wave state (potentially with a symmetry-allowed triplet admixture), is the most common $M_z$-even state favored by electron-phonon interactions, recent works show that strong spin fluctuations in this system should favor the $E'$ channel \cite{Roy24,Siegl24}. 

In addition to being unaffected by Ising SOC and favored by spin fluctuations, the $E'$ channel has a number of appealing features. First, as a two-component order parameter it admits both chiral and nematic ground states, in line with the phenomenology in 4H$_b$-TaS$_2$\cite{Ribak20,Persky22,Nayak21,Almoalem24,Silber24}. In addition, this state has interesting topological features. $M_z$-even pairing with Ising SOC implies spin rotation invariance around $z$ direction, which puts this system in topological class AIII \cite{Schnyder08}, featuring $Z$ invariants in 1D that can protect nodal points in the case of nematic ground states. For chiral ground states, each mirror sector realizes a Chern number $C=3$ superconductor. In the case of triplet $E'$ chiral state, a $C=6$ superconductor is realized, which could also be consistent with the edge state phenomenology in 4H$_b$-TaS$_2$\cite{Nayak21} and CrBr$_3$/NbSe$_2$\cite{Kezilebieke20}.

\begin{figure*}[h!t]
    \includegraphics[width=\linewidth]{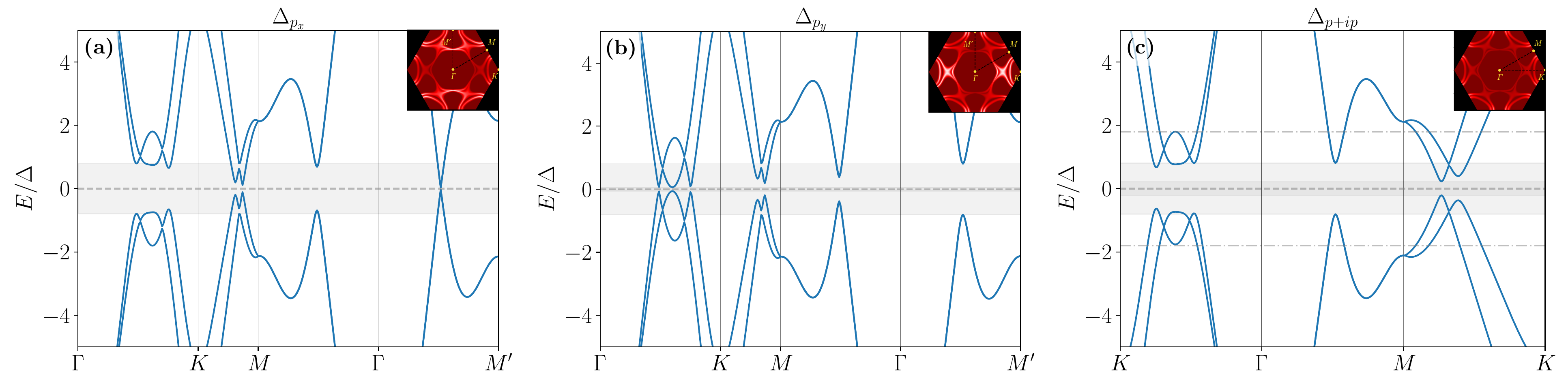}
    \caption{Band structure of the BdG Hamiltonian in Eq.~\eqref{eqn:bdg_hamiltonian} for 4Hb-TaS$_2$ with the superconducting pairings (a) $\Delta_{p_x}$, (b) $\Delta_{p_y}$, and (c) $\Delta_{p+ip}$ for $\Delta=0.1$ eV. The dark (light) gray shading marks the minimal (maximal) gap along high-symmetry directions. Insets show the gap magnitude across the Brillouin zone; light colors indicate smaller gaps and insets mark the paths used in the main panels. The dotted-dashed line in (c) marks the energy associated with the van Hove singularity that produces the secondary peak in the absorptive components $\mathrm{Re}\sigma_{xx}$ and $\mathrm{Im}\sigma^H_{xy}$.}
    \label{fig:band_structure_delta_px_py_pip_low_energy}
\end{figure*}

The discussion above motivates us to consider the triplet $E'$ channel, given in the orbital basis $\{d_{z^2},d_{xy},d_{x^2-y^2}\}$ as the two-component basis pairing matrices

\begin{equation}
\{\Delta_{p_x},\Delta_{p_y}\} =     \sigma_y\sigma_z\otimes\left\lbrace{\left(\begin{array}{ccc}
0 & i & 0 \\
-i & 0 & 0 \\
0 & 0 & 0
\end{array}\right)}, {\left(\begin{array}{ccc}
0 & 0 & -i \\
0 & 0 & 0 \\
i & 0 & 0
\end{array}\right)}\right\rbrace.
\end{equation} 

This term represents spin-triplet, orbital singlet pairing between the out-of-plane orbital $d_{z^2}$ and each of the in-plane orbitals $d_{x^2-y^2}$ and $d_{xy}$. The superconducting phases with pairings $\Delta_{p_x}$ and $\Delta_{p_y}$ represent nematic phases, while the chiral superconducting phase is given by the pairing $\Delta_{p+ip}=\Delta_{p_x}+i\Delta_{p_y}$. 

\section{BdG quasiparticle band structure}

For concreteness, we consider a single layer of H-TaS$_2$ and use the normal state tight-binding parameters reported in Ref.~[\onlinecite{Margalit21}]. \edits{Band structure effects and optical features of the superconducting state with realistic values of the superconducting gap ($\sim$1 meV) are, in principle, quite small. For illustrative purposes we consider pairings of an artificially large strength $\Delta=0.1$ eV, keeping in mind that experimental features are to be expected at the real values of the superconducting gap. We will focus on realistic values of $\Delta$ and the scaling of the relevant features later on}. The three different ground states with $E'$ pairing $\Delta_{p_x,p_y,p+ip}$ lead to the three different band structures in Fig.~\ref{fig:band_structure_delta_px_py_pip_low_energy}. The $\Delta_{p_x}$ pairing is gapless along the $\Gamma M'$ ($k_x=0$) direction, while the $\Delta_{p_y}$ pairing is gapless along the $\Gamma K$ ($k_y=0$) direction. In both cases we see that $\Gamma M$ and $\Gamma M'$ are not equivalent, a consequence of the breaking of $C_3$ symmetry. For both nematic phases, the maximum gap is approximately $1.6\Delta$ along the $\Gamma K$ ($\Gamma M'$) directions for $\Delta_{p_x}$ ($\Delta_{p_y}$). The $\Delta_{p+ip}$ phase, on the other hand, restores $C_3$ symmetry (combined with a gauge transformation) so the energies are $C_3$-symmetric, and it exhibits a full gap ranging in size from $0.43\Delta$ along the $MK$ direction to $1.6\Delta$ along the $\Gamma M$ direction.

\section{Optical conductivity} 
\FJ{
To compute the optical conductivity, defined from  $J_i = \sigma_{ij} E_j$, we employ the standard Kubo formalism as used for superconductors in Ref.~[\onlinecite{Ahn21}]. Since we consider $k-$independent pairings, Peierls substitution couples the electromagnetic field $\vec k \to \vec k - e \vec A$ through the normal-state Hamiltonian $h_k$. The conductivity tensor $\sigma_{ij}$ can be separated into symmetric and antisymmetric (Hall) parts. For the symmetric part we only consider the diagonal components. In the main text we report the absorptive part of the conductivity $\sigma_{ij}^{\rm abs}=\tfrac{1}{2}(\sigma_{ij}+\sigma_{ji}^*)$. For the symmetric part, the absorptive part is real $\sigma_{ii}^{\rm abs} = {\rm Re}\sigma_{ii}$, while for the Hall part $\sigma^{H}_{xy}=\frac{1}{2}\left(\sigma_{xy}-\sigma_{yx}\right)$, the absorptive part is imaginary $\sigma_{ij}^{H,\rm abs} = {\rm Im}\sigma_{ij}^H$. The reactive parts $\sigma_{ij}^{\rm rea}=\tfrac{1}{2}(\sigma_{ij}-\sigma_{ji}^*)$ can be computed through Kramers-Kronig relations (see Appendices~\ref{appendix:realimagoptcond} and ~\ref{appendix:kramers_kronig}). 
The absorptive, diagonal part is explicitly given by
\begin{align}
\label{eqn:realoptcond}
    {\rm Re}\sigma_{ii} = \frac{\pi e^2}{2 \omega V} \sum_{m \neq n} v^i_{nm} v^i_{mn} f_{nm} \delta(\varepsilon_n-\varepsilon_m + \omega),
\end{align}
while the absorptive Hall part is given by 
\begin{align}
\label{eqn:hallcond}
   {\rm Im} \sigma^H_{xy} 
    =\frac{\pi e^2}{2\omega V}\sum_{m\neq n}\mathrm{Im}\left[v_{nm}^x v_{mn}^y\right]f_{nm}\delta\left(\epsilon_n-\epsilon_m +\omega\right),
\end{align}}
where $e$ is the electron charge, $V$ is the volume of the system, $f_{nm}=f(\varepsilon_n)-f(\varepsilon_m)$ is the difference of Fermi functions, which depend on the energy $\varepsilon_n= E_n -\mu$ of band $n$, the chemical potential $\mu$ and the inverse temperature $\beta=1/k_B T$ where $k_B$ is the Boltzmann constant. We set $\beta=300\mathrm{eV}^{-1}$ ($T\simeq 3.87$K). The velocity matrix elements $v^i_{mn}$ account for the transition amplitude between states $n,m$ and are defined as:
\begin{align}
    v^i_{nm}=\bra{n}\frac{\partial H}{\partial k_i} \ket{m} = \bra{n} 
    \left(\begin{array}{cc}
\partial_{k_i} h_k & 0 \\
0 & \partial_{k_i}\left[h^T_{-k}\right]
\end{array}\right) 
\ket{m}.
\end{align}

Note that the optical conductivity of clean multiband superconductors\cite{Ahn21} as computed here can be reduced in size for some types of superconductors—but not suppressed—by the inclusion of vertex corrections\cite{watanabe_gauge-invariant_2024}, which are outside the scope of this work. 

\section{Anisotropy in the nematic phase}

\begin{figure}[t]
    \centering
    \includegraphics[width=\linewidth]{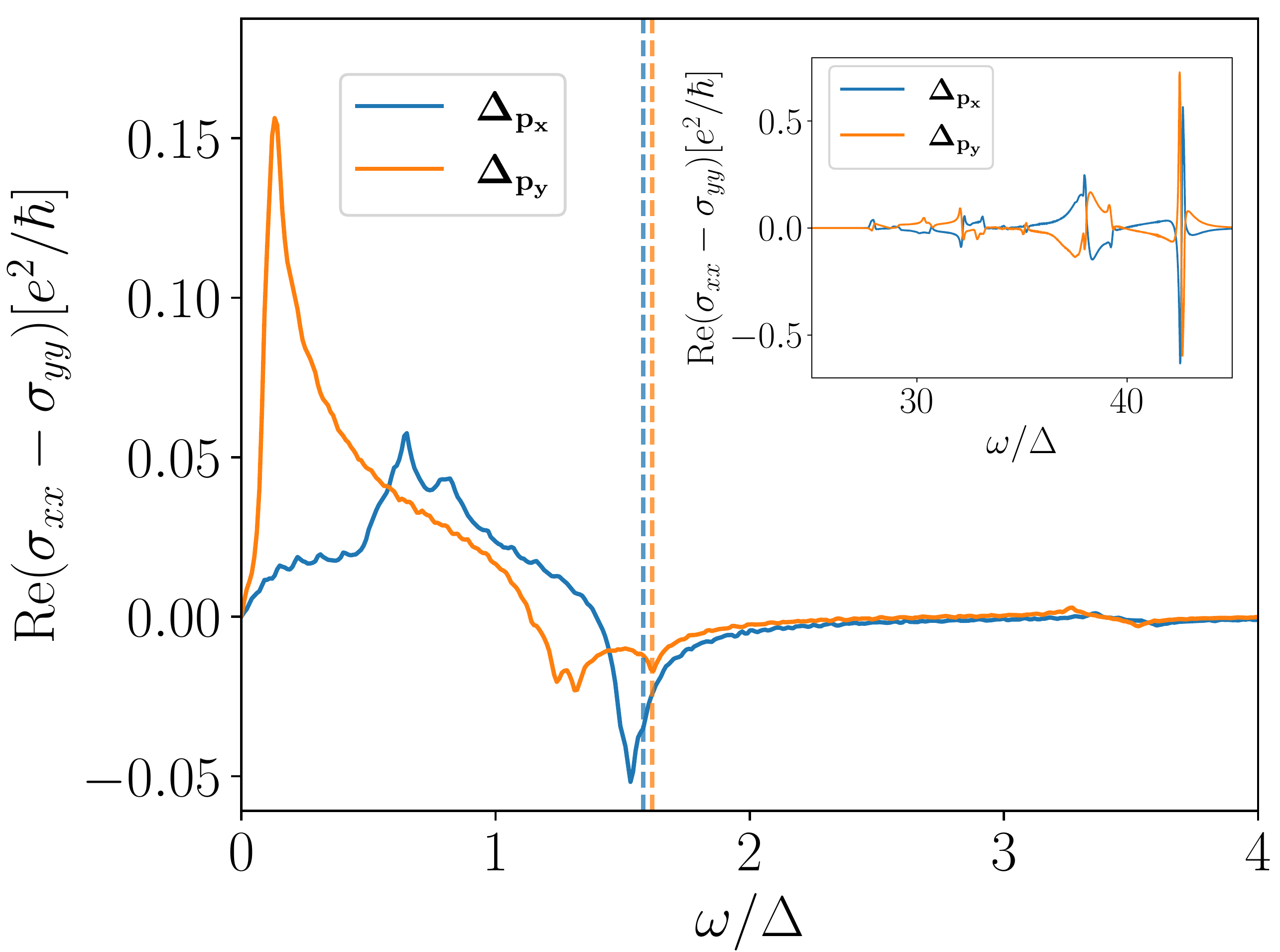}
    \caption{Anisotropy of the absorptive part of the optical conductivity at low (main figures) and high (inset) frequencies for the nematic phases $\Delta_{p_x}$ and $\Delta_{p_y}$ with $\Delta=0.1$ eV.}
    \label{fig:anisotrpy_nematic_phases}
\end{figure}

In the presence of $C_3$ symmetry, the optical conductivity satisfies $\sigma_{xx} = \sigma_{yy}$. Since the two-component nematic pairing breaks this symmetry, we consider the difference $\sigma_{xx} - \sigma_{yy}$ as a probe of the nematic pairing. Fig.~\ref{fig:anisotrpy_nematic_phases} shows this difference in the range of frequencies near the superconducting gap for both $\Delta_{p_x}$ and $\Delta_{p_y}$. A sizable conductivity occurs due to transitions near the Fermi level, with significant differences between $\Delta_{p_x}$ and $\Delta_{p_y}$ that can help determine the pairing anisotropy.

As a further optical signature of the pairing in multiorbital systems, we also consider higher frequency effects that occur due to transitions from the Fermi level BdG quasiparticles to unoccupied electron states, and from occupied electron states to BdG quasiparticles. These occur in the frequency range where the optical conductivity of the normal state is finite (see Fig.~\ref{fig:Hlayernormalstate} (c)) and represent the transfer of spectral weight due to superconductivity as discussed in Ref.~[\onlinecite{Ahn21a}]. However, here we consider responses which are zero in the absence of pairing, due to $C_3$ and time-reversal symmetry. The observation of any optical conductivity is therefore a signature of multiorbital pairing, which is shown in the inset of Fig.~\ref{fig:anisotrpy_nematic_phases}. The anisotropy of the nematic pairing is reflected in this spectrum as well. 

\section{Hall conductivity in the chiral phase\label{sec:hall_conductivity_chiral}}

\begin{figure}[h!]
    \centering
    \includegraphics[width=\linewidth]{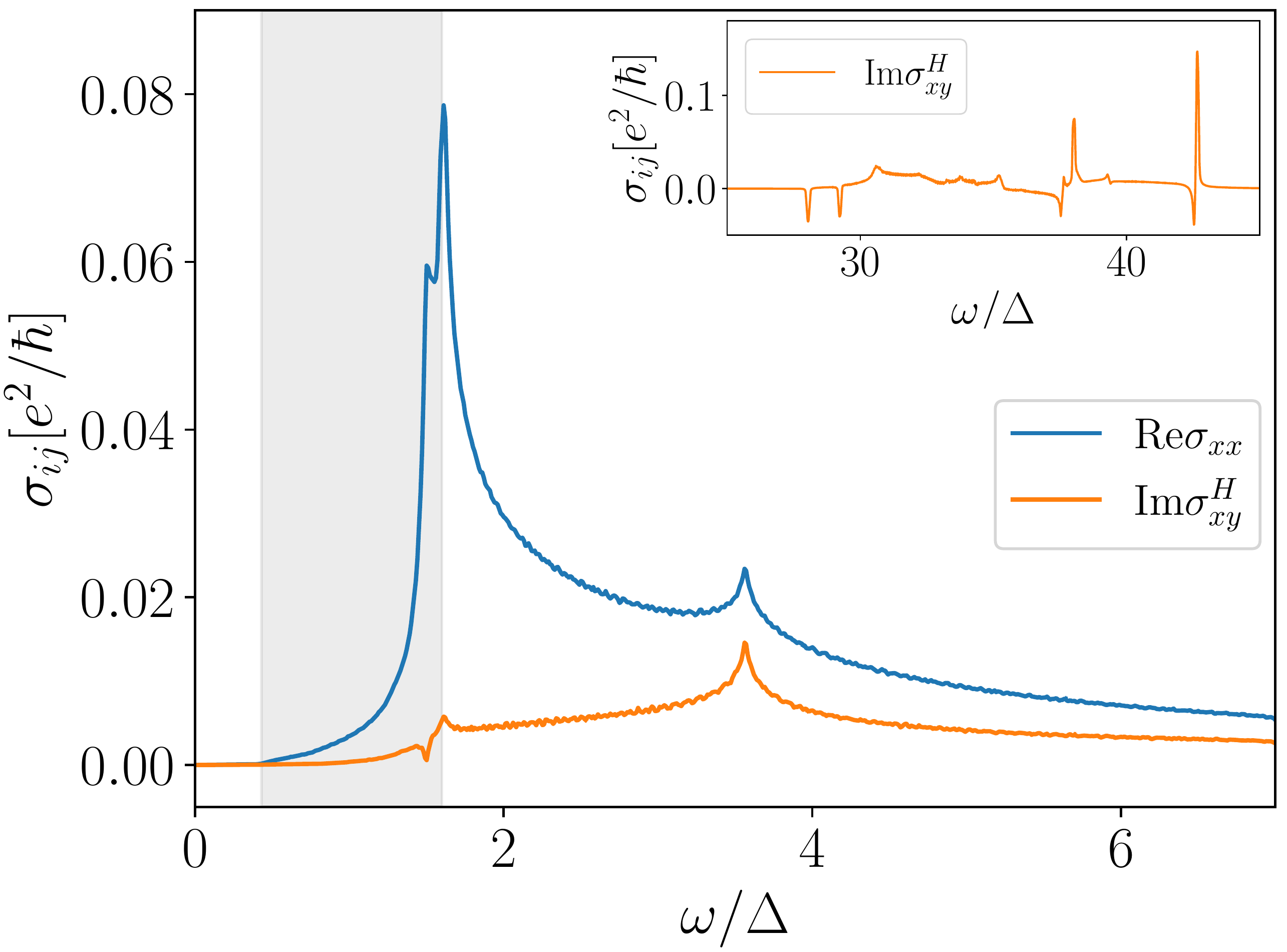}
    \caption{\edits{Hall response in the chiral state. We compare the anti-symmetric and symmetric absorptive components of the optical conducitivty, $\mathrm{Im}\sigma^H_{xy}$ and ${\rm Re}\sigma_{xx}$, respectively, for $\Delta_{p+ip}$ at $\Delta=0.1$ eV at low (main) and high (inset) frequencies. The shadowed region indicates the span of the gap across the Brillouin Zone.}}
    \label{fig:hall_conductivity_pip}
\end{figure}

\begin{figure}
    \centering
    \includegraphics[width=\linewidth]{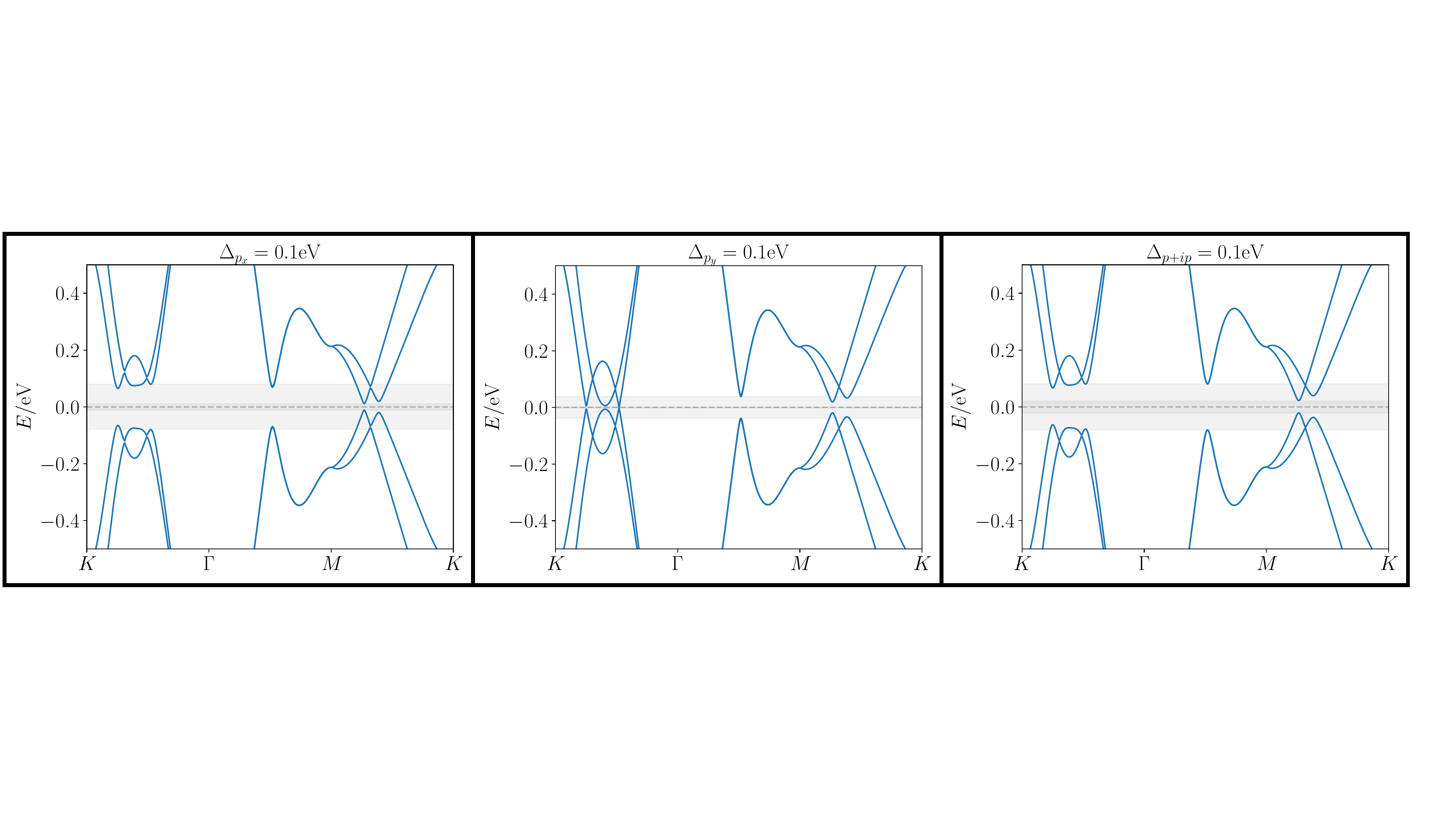}
    \caption{Brillouin Zone-resolved absorptive part of the optical conductivity ${\rm Re}\sigma_{xx}$ (left column) and ${\rm Im}\sigma^H_{xy}$ (right) column integrated between $\omega=1.3\Delta$ and $\omega=1.8\Delta$ (upper row) corresponding to the low-energy peak in the response in Fig.~\ref{fig:hall_conductivity_pip}, and $\omega=3\Delta$ and $\omega=4\Delta$ (lower row) corresponding to the secondary peak in the response in Fig.~\ref{fig:hall_conductivity_pip}.}
    \label{fig:bz_optical_response}
\end{figure}

\edits{The chiral $E'$ state ($\Delta_{p+ip}$) breaks TRS, enabling a finite optical Hall response $\sigma_{xy}^H$. Figure~\ref{fig:hall_conductivity_pip} compares the absorptive components $\mathrm{Re}\sigma_{xx}$ and $\mathrm{Im}\sigma_{xy}^H$ across the gap window. The $\mathrm{Re}\sigma_{xx}$ component exhibits a large peak around $1.6\Delta$ associated with the large JDOS between the closest bands to the Fermi level along the $K\Gamma$ high-symmetry lines and the positive-defined optical conductivity integrand over the Brillouin Zone (see Fig.~\ref{fig:bz_optical_response} (a)). However, in the case of $\mathrm{Im}\sigma^H_{xy}$ the velocity matrix elements connecting the closest bands to the Fermi level over the Brillouin Zone can be positive or negative (see Fig.~\ref{fig:bz_optical_response} (b)). This leads to a cancellation of contributions that suppresses the peak that is present in the $\mathrm{Re}\sigma_{xx}$ component due to the positive value of its velocity matrix elements over the whole Brillouin Zone.}

\edits{The secondary peak at $\omega\!\sim\!3.6\Delta$, associated with a higher‑frequency van Hove singularity along $K\Gamma$ [Fig.~\ref{fig:band_structure_delta_px_py_pip_low_energy}(c)], is similar in size for $\sigma_{xx}$ and $\sigma^H_{xy}$ [Fig.~\ref{fig:hall_conductivity_pip}]. Here positive velocity‑matrix contributions dominate, resembling the longitudinal case [Figs.~\ref{fig:bz_optical_response}(c),(d)]. The predominance arises from the large positive weight of the van Hove feature along $K\Gamma$ and the reduced presence of negative contributions in this window [Fig.~\ref{fig:bz_optical_response}(d)], in contrast to the lower‑energy peak [Fig.~\ref{fig:bz_optical_response}(b)].}

Additionally, the $\Delta_{p+ip}$ also exhibits a finite contribution to the Hall conductivity at large frequencies due to the electronic transitions from Fermi surface BdG quasiparticles to higher-energy unpaired bands, shown in the inset of Fig.~\ref{fig:hall_conductivity_pip}. This provides an additional high-energy fingerprint of the TRS breaking associated exclusively with the chiral phase.

\newpage
\textcolor{black}{\section{Experimental outlook}}

After establishing the general features of the band structure and optical conductivity for large values of $\Delta$ in the previous sections, we now focus on the measurability of the optical properties in the nematic and chiral phases. To do so, we will consider a realistic value of the superconducting order parameter $\Delta=1$ meV.\\

\subsection{Nematic phases}

\begin{figure}
    \centering
    \includegraphics[width=\linewidth]{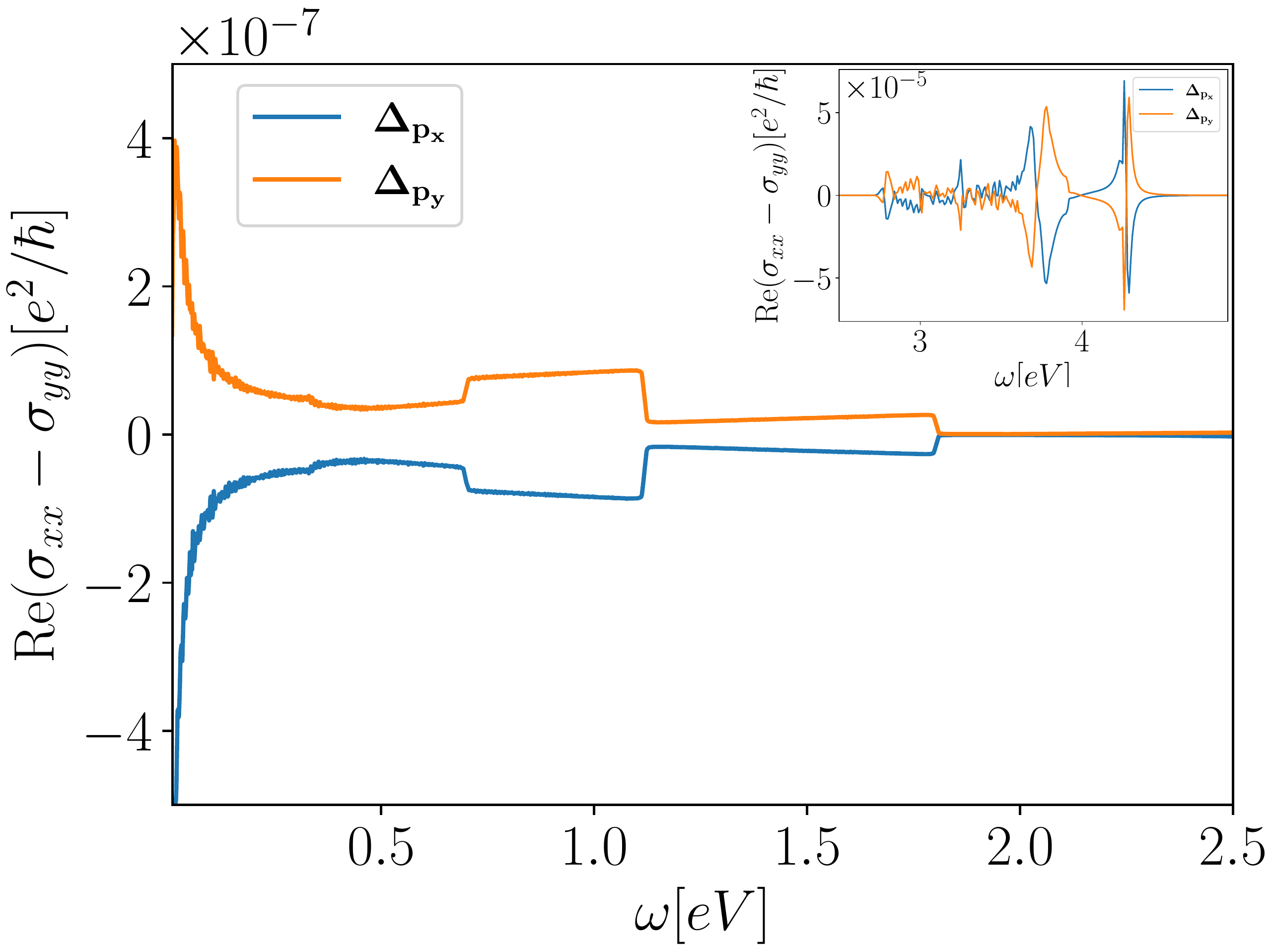}
    \caption{Anisotropy of the absorptive components of the optical conductivity $\mathrm{Re}(\sigma_{xx}-\sigma_{yy})$ of the nematic superconducting states $\Delta_{p_x}$ and $\Delta_{p_y}$ for a realistic $\Delta=1$~meV in the $\omega\in[0,2.5]$~eV (main) and $\omega\in[2.5,4.5]$~eV (inset) regions.}
    \label{fig:anisotropy_001eV}
\end{figure}

In the nematic phases $\Delta_{p_x}$ and $\Delta_{p_y}$, the small gap $\Delta=1$ meV results in two well-differentiated $\omega$ regions with different anisotropy, as we see in Fig.~\ref{fig:anisotropy_001eV}. The first region, spanning from $\omega=0$~eV to around $\omega=2$~eV features an anisotropy of the absorptive components of the optical conductivity $\mathrm{Re}(\sigma_{xx}-\sigma_{yy})\sim 10^{-7} e^2/\hbar$. The relevant quantity to consider experimental feasibility is the relative size of the anisotropy compared to the optical conductivity in that region, $\Delta\sigma/\sigma\equiv\frac{\mathrm{Re}(\sigma_{xx}-\sigma_{yy})}{\mathrm{Re}(\sigma_{xx}+\sigma_{yy})/2}\sim 10^{-6}$. 

The second region (Fig.~\ref{fig:anisotropy_001eV}, inset) spans from $\omega=2.5$~eV to around $\omega=4.5$~eV. The anisotropy here is more than two orders of magnitude larger than that in the lower-energy region described in the previous paragraph, reaching values of $\mathrm{Re}(\sigma_{xx}-\sigma_{yy})\sim 5\times 10^{-5} e^2/\hbar$. The average value of the optical conductivity in this region is barely affected by the nematic phase, and its size is similar to that of the normal state in Fig.~\ref{fig:Hlayernormalstate} (c). Therefore, the relative size of the anisotropy in this region is $\Delta\sigma/\sigma\sim 10^{-5}$. 

The experimental feasibility of detecting optical anisotropies of magnitude $\Delta\sigma/\sigma\sim 10^{-6}$--$10^{-5}$ is well-supported by current instrumentation standards\cite{harrison2018development,PhysRevLett.121.027001}. In the specific context of rotational symmetry breaking, reflectance anisotropy measurements have resolved small reflectivity features with a relative size of $\Delta R/R \sim 10^{-6}$ to characterize anisotropy in nematic superconductors\cite{PhysRevLett.121.027001}. Therefore, the expected anisotropy signature of the nematic superconducting phases discussed here for a realistic $\Delta=1$~meV is experimentally accessible with the current techniques.



\subsection{Chiral phase}



The chiral $E'$ phase produces a finite optical Hall conductivity $\sigma_{xy}^H$, which is most directly accessed via the polar Kerr effect (rotation of the reflected polarization) measured in a normal-incidence setup~\cite{argyres1955theory,Kapitulnik2009}. The Kerr angle is given by\cite{argyres1955theory,Kapitulnik2009}

\begin{equation}
    \label{eqn:kerr_angle}
    \theta_{\mathrm{K}}=-\mathcal{I} m\left[\frac{\tilde{n}_{\mathrm{L}}-\tilde{n}_{\mathrm{R}}}{\tilde{n}_{\mathrm{L}} \tilde{n}_{\mathrm{R}}-1}\right],
\end{equation}
where $\tilde{n}\equiv n + i\kappa$ is the complex refractive index for left $\tilde{n}_{\mathrm{L}}$ and right $\tilde{n}_{\mathrm{R}}$ circularly polarized light. In the case of a 2D sample with no substrate, the Kerr angle can be approximated by\cite{Kapitulnik2009,PhysRevLett.105.057401}

\begin{align}
    \theta_K\approx\tan\theta_K = \mathrm{Re}\left[\frac{\sigma_{xy}^H}{\sigma_{xx}}\right],
\end{align}
where $\sigma_{xx}$ and $\sigma_{xy}^H$ are the complex conductivities, and we have used the fact that $\theta_K\ll1$ in $\tan\theta_K\approx\theta_K$.







\begin{figure}[h!]
    \centering
    \includegraphics[width=\linewidth]{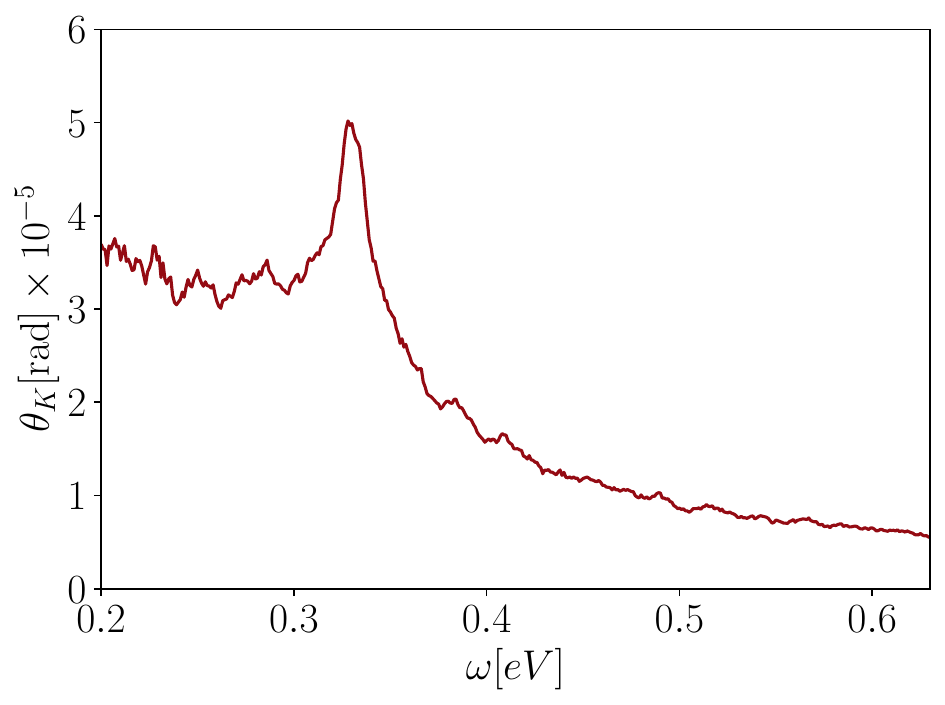}
    \caption{Kerr angle $\theta_K(\omega)$ for TaS$_2$ with chiral $p+ip$ pairing with $\Delta=1$~meV. For $\omega<0.2$~eV the small size of $\Delta$ results in a large numerical noise and an absence of distinguishable features.
    }
    \label{fig:kerr_angle_monolayer}
\end{figure}

The Kerr angle for TaS$_2$ with $\Delta=1$ meV in the chiral superconducting phase is shown in Fig.~\ref{fig:kerr_angle_monolayer}. Remarkably, the Kerr angle in this setup is in the $10^{-5}$--$10^{-4}$ radians range, considerably larger than the resolution limit of $10^{-9}$ radians measured in other setups using Sagnac interferometry\cite{Kapitulnik2006,Kapitulnik2009}. 

The large size of the Kerr rotation angle is due to the simultaneous scaling of both components, $\sigma_{xy}$ and $\sigma_{xx}$, with $\Delta$ in the described range. For values of $\Delta\in[1,10]$ meV we observe a strong peak at $\omega=330$ meV for all $\Delta$ values considered, indicating an interplay between superconductivity and the intrinsic features of the band structure that do not depend on the size of $\Delta$, leading to an observable feature in the Kerr angle. This is the same feature that gets displaced to $\omega=360$ meV for $\Delta=100$ meV, as discussed in Figs.~\ref{fig:hall_conductivity_pip} and \ref{fig:bz_optical_response}, due to the large value of $\Delta$. This means that the cause of the persistent peak at $\omega=330$meV  in Fig.~\ref{fig:kerr_angle_monolayer} is the same physics discussed in Sec.~\ref{sec:hall_conductivity_chiral}.

\begin{figure*}
    \centering
    \includegraphics[width=\linewidth]{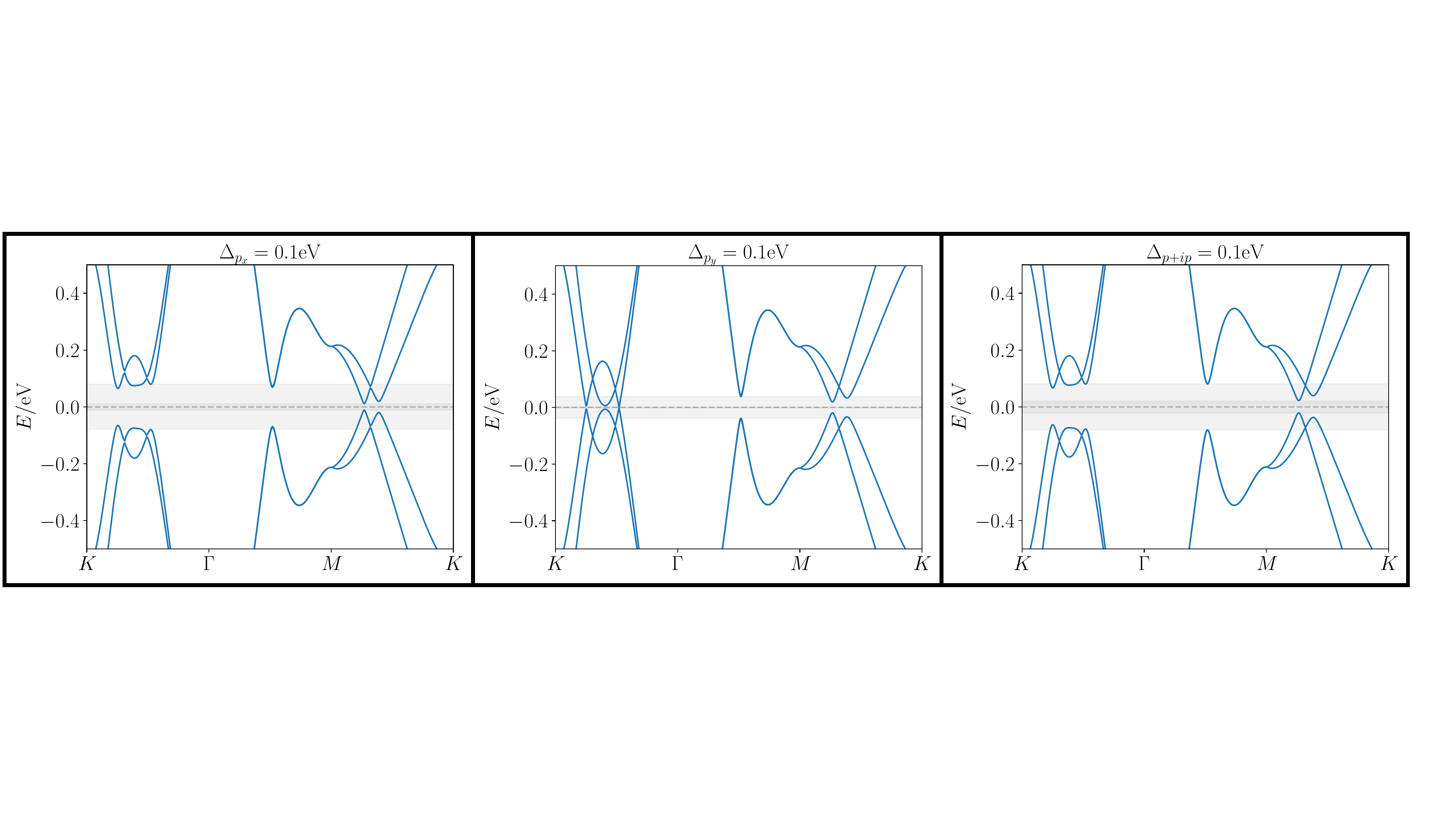}
    \caption{\edits{
    Persistent features and scaling of the Kerr angle for $\Delta\in[1,10]$meV. (a)
    Kerr angle $\theta_K(\omega)$ for a single H-layer of TaS$_2$ with chiral $p+ip$ pairing for different values of $\Delta$. The position of the persistent peak at $\omega=330$ meV is indicated by the dashed gray line. For $\omega<0.2$~eV there is a strong numerical noise background covering any discernible features. (b) Scaling of the integrated Kerr angle $\theta_K\sim \Delta^b$ in the range $\omega\in[0,0.7]$ for different values of $\Delta$. The least squares fit of the shown data shows a quadratic exponent.}
    }
    \label{fig:kerr_angle_scaling}
\end{figure*}

Finally, we see that the size of the Kerr rotation angle in the Kerr-active region scales quadratically with $\Delta$ over a range of $1-10$ meV (see Fig.~\ref{fig:kerr_angle_scaling}), providing an estimation of the order of magnitude of the angle for setups with similar properties and different superconducting pairings.

\phantom{a}

\section{Discussion}

In our work, we have predicted the optical conductivity of single layers of an H polytype of a TMD with $D_{3h}$ symmetry. These results most directly apply to single layers of NbSe$_2$, where signatures of a nematic pairing channel have been observed in Refs.~[\onlinecite{Hamill21}] and [\onlinecite{Cho22}]. Our results also bear relevance to 4Hb-TaS$_2$, where superconductivity is dominated by the H layers in the T/H bulk structure. It should be noted this structure has an extra inversion center between the H layers and there are therefore two possibilities for inversion-even and odd versions of the $E'$ state, which correspond to irreps $E_{2g}$ and $E_{1u}$ of the group $D_{6h}$ and should remain nearly degenerate \DM{if the pairing is mainly intralayer}. Finally, we speculate that our results may also be relevant to the superconducting state observed in CrBr$_3$/NbSe$_2$\cite{Kezilebieke20}, which coexists with ferromagnetism and shows chiral edge modes corresponding to a topological chiral superconductor.

\edits{Our quantitative outlook indicates that chiral $E'$ order in H-TaS$_2$ should produce Kerr rotations of $\theta_K\!\sim\!10^{-5}$ rad for realistic gap sizes, well within the sensitivity of modern Sagnac interferometry\cite{Kapitulnik2006,Kapitulnik2009}. 
Additionally, the nematic $E'$ state yields a relative diagonal anisotropy $\mathrm{Re}(\sigma_{xx}-\sigma_{yy})/\sigma_{xx}\sim10^{-5}$ without a Hall response, well within current experimental limits, enabling an optical symmetry diagnostic using polarization-resolved reflectivity. Likely, the experimental feasibility of the optical responses discussed here extends to other TMDs with superconducting gaps in the order of meV.}



\edits{In summary, our results chart a concrete experimental pathway to help establish whether unconventional superconductivity in different TMD platforms is due to $E'$ pairing: use linear optics to distinguish nematic versus chiral $E'$ order, and quantify chiral TRSB via Kerr rotation.}

\section{Data and code availability}
\edits{The code used to compute the optical conductivity and the generated datasets supporting this work will be made publicly available via Zenodo upon publication; the DOI will be inserted in the final version.}

\section{Acknowledgements}

F. J. is supported by Grant PID2021-128760NB0-I00 from the Spanish MCIN/AEI/10.13039/501100011033/FEDER, EU. M.-Á.S.-M. was supported by EPSRC grant EP/X012239/1 and Programa Red Guipuzcoana de Ciencia Tecnología e Innovación No. 2021-CIEN-000070-01 Gipuzkoa Next. D.M-S. acknowledges financial support from the Spanish Ministerio de Ciencia, Innovaci\'on y Universidades (MCIU) FPU fellowship No. FPU19/03195, and from the National Science Foundation (NSF) Materials Research Science and Engineering Centers (MRSEC) program through Columbia University under the Precision-Assembled Quantum Materials (PAQM) Grant No. DMR-2011738.


\appendix
\textcolor{black}{\section{Real and imaginary parts of the optical conductivity \label{appendix:realimagoptcond}}}

The complete expression for the interband optical conductivity, containing real and imaginary parts, can be derived using standard Kubo formalism and reads 

\begin{eqnarray}
\label{eqn:compleoptcond}
\sigma_{ij} &=& \frac{\pi e^2}{2 \omega V} \sum_{m \neq n} \frac{v^i_{nm} v^j_{mn} f_{nm}}{\varepsilon_n-\varepsilon_m + \omega+i\delta}\\
&\underset{\delta\to0}{\to}& \frac{\pi e^2}{2 \omega V} \sum_{m \neq n} \left[\mathcal{P} -i\pi\delta(\varepsilon_n-\varepsilon_m + \omega) \right] \\
&&\times \left[\mathrm{Re}(v^i_{nm} v^j_{mn})+i\mathrm{Im}(v^i_{nm} v^j_{mn})\right] f_{nm},
\end{eqnarray}
where we have used the Dirac identity 

\begin{eqnarray}
    \frac{1}{z+i\delta}\underset{\delta \to 0}{\to} \mathcal{P}\int_{-\infty}^{\infty}\mathrm{d}z\frac{1}{z}-i\pi\delta(z),
\end{eqnarray}
where $\mathcal{P}$ is the Principal value of the integral and $\delta(z)$ is the Dirac delta function.

Taking the real part of Eq.~\eqref{eqn:compleoptcond} and noting that $v_{nm}^i v_{mn}^i = |v_{nm}^i|^2$ is a real quantity, we obtain the expression in Eq.~\eqref{eqn:realoptcond} for the real part of the diagonal components of the optical conductivity. 

Likewise, taking the imaginary part of Eq.~\eqref{eqn:compleoptcond} for the antisymmetric, off-diagonal terms $i\neq j$ we obtain the expression for the Hall conductivity in Eq.~\eqref{eqn:hallcond}.

\textcolor{black}{\section{Kramers-Kronig relations\label{appendix:kramers_kronig}}}

The real and imaginary parts of the optical conductivity tensor $\mathrm{Re}\sigma + i\mathrm{Im}\sigma$ are related by the Kramers-Kronig identities: 
\begin{eqnarray}
\label{eqn:KKPvalue}
\mathrm{Re}\sigma(\omega)&=&\frac{1}{\pi}\mathcal{P}\int_{-\infty}^{\infty}\mathrm{d}x\frac{\mathrm{Im}\sigma(x)}{x-\omega},\\
\mathrm{Im}\sigma(\omega)&=&-\frac{1}{\pi}\mathcal{P}\int_{-\infty}^{\infty}\mathrm{d}x\frac{\mathrm{Re}\sigma(x)}{x-\omega}.
\end{eqnarray}

Noting that $\mathrm{Re}\sigma(\omega)$ is even and $\mathrm{Im}\sigma(\omega)$ is odd in frequency, and following the standard derivation~\cite{dresselhaus_optical_nodate}, we obtain
\begin{eqnarray}
    \label{eqn:KK}
    \mathrm{Re}\sigma(\omega)&=&\frac{2}{\pi}\int_{0}^{\infty}\mathrm{d}x\frac{x\mathrm{Im}\sigma(x)-\omega\mathrm{Im}\sigma(\omega)}{x^2-\omega^2},\\
    \mathrm{Im}\sigma(\omega)&=&-\frac{2\omega}{\pi}\int_0^{\infty}\mathrm{d}x\frac{\mathrm{Re}\sigma(x)-\mathrm{Re}\sigma(\omega)}{x^2-\omega^2}.
\end{eqnarray}

Since the bandwidth of the model used in the main text is finite, the absorptive components of $\sigma(\omega)$ computed in the main text are supported on a bounded frequency window and the integrals in Eq.~\eqref{eqn:KK} converge without ad-hoc truncations. To ensure convergence and numerical stability, the integrals in Eqs.~\eqref{eqn:KK} are computed for a maximum $\omega_{\mathrm{cutoff}}=50$eV.

\newpage

\begin{thebibliography}{59}%
\makeatletter
\providecommand \@ifxundefined [1]{%
 \@ifx{#1\undefined}
}%
\providecommand \@ifnum [1]{%
 \ifnum #1\expandafter \@firstoftwo
 \else \expandafter \@secondoftwo
 \fi
}%
\providecommand \@ifx [1]{%
 \ifx #1\expandafter \@firstoftwo
 \else \expandafter \@secondoftwo
 \fi
}%
\providecommand \natexlab [1]{#1}%
\providecommand \enquote  [1]{``#1''}%
\providecommand \bibnamefont  [1]{#1}%
\providecommand \bibfnamefont [1]{#1}%
\providecommand \citenamefont [1]{#1}%
\providecommand \href@noop [0]{\@secondoftwo}%
\providecommand \href [0]{\begingroup \@sanitize@url \@href}%
\providecommand \@href[1]{\@@startlink{#1}\@@href}%
\providecommand \@@href[1]{\endgroup#1\@@endlink}%
\providecommand \@sanitize@url [0]{\catcode `\\12\catcode `\$12\catcode `\&12\catcode `\#12\catcode `\^12\catcode `\_12\catcode `\%12\relax}%
\providecommand \@@startlink[1]{}%
\providecommand \@@endlink[0]{}%
\providecommand \url  [0]{\begingroup\@sanitize@url \@url }%
\providecommand \@url [1]{\endgroup\@href {#1}{\urlprefix }}%
\providecommand \urlprefix  [0]{URL }%
\providecommand \Eprint [0]{\href }%
\providecommand \doibase [0]{http://dx.doi.org/}%
\providecommand \selectlanguage [0]{\@gobble}%
\providecommand \bibinfo  [0]{\@secondoftwo}%
\providecommand \bibfield  [0]{\@secondoftwo}%
\providecommand \translation [1]{[#1]}%
\providecommand \BibitemOpen [0]{}%
\providecommand \bibitemStop [0]{}%
\providecommand \bibitemNoStop [0]{.\EOS\space}%
\providecommand \EOS [0]{\spacefactor3000\relax}%
\providecommand \BibitemShut  [1]{\csname bibitem#1\endcsname}%
\let\auto@bib@innerbib\@empty
\bibitem [{\citenamefont {Wilson}\ \emph {et~al.}(1975)\citenamefont {Wilson}, \citenamefont {Di~Salvo},\ and\ \citenamefont {Mahajan}}]{Wilson75}%
  \BibitemOpen
  \bibfield  {author} {\bibinfo {author} {\bibfnamefont {J.~A.}\ \bibnamefont {Wilson}}, \bibinfo {author} {\bibfnamefont {F.}~\bibnamefont {Di~Salvo}}, \ and\ \bibinfo {author} {\bibfnamefont {S.}~\bibnamefont {Mahajan}},\ }\href@noop {} {\bibfield  {journal} {\bibinfo  {journal} {Advances in Physics}\ }\textbf {\bibinfo {volume} {24}},\ \bibinfo {pages} {117} (\bibinfo {year} {1975})}\BibitemShut {NoStop}%
\bibitem [{\citenamefont {Norman}(2011)}]{Norman11}%
  \BibitemOpen
  \bibfield  {author} {\bibinfo {author} {\bibfnamefont {M.~R.}\ \bibnamefont {Norman}},\ }\href@noop {} {\bibfield  {journal} {\bibinfo  {journal} {Science}\ }\textbf {\bibinfo {volume} {332}},\ \bibinfo {pages} {196} (\bibinfo {year} {2011})}\BibitemShut {NoStop}%
\bibitem [{\citenamefont {Hamill}\ \emph {et~al.}(2021)\citenamefont {Hamill}, \citenamefont {Heischmidt}, \citenamefont {Sohn}, \citenamefont {Shaffer}, \citenamefont {Tsai}, \citenamefont {Zhang}, \citenamefont {Xi}, \citenamefont {Suslov}, \citenamefont {Berger}, \citenamefont {Forr{\'o}} \emph {et~al.}}]{Hamill21}%
  \BibitemOpen
  \bibfield  {author} {\bibinfo {author} {\bibfnamefont {A.}~\bibnamefont {Hamill}}, \bibinfo {author} {\bibfnamefont {B.}~\bibnamefont {Heischmidt}}, \bibinfo {author} {\bibfnamefont {E.}~\bibnamefont {Sohn}}, \bibinfo {author} {\bibfnamefont {D.}~\bibnamefont {Shaffer}}, \bibinfo {author} {\bibfnamefont {K.-T.}\ \bibnamefont {Tsai}}, \bibinfo {author} {\bibfnamefont {X.}~\bibnamefont {Zhang}}, \bibinfo {author} {\bibfnamefont {X.}~\bibnamefont {Xi}}, \bibinfo {author} {\bibfnamefont {A.}~\bibnamefont {Suslov}}, \bibinfo {author} {\bibfnamefont {H.}~\bibnamefont {Berger}}, \bibinfo {author} {\bibfnamefont {L.}~\bibnamefont {Forr{\'o}}},  \emph {et~al.},\ }\href@noop {} {\bibfield  {journal} {\bibinfo  {journal} {Nature physics}\ }\textbf {\bibinfo {volume} {17}},\ \bibinfo {pages} {949} (\bibinfo {year} {2021})}\BibitemShut {NoStop}%
\bibitem [{\citenamefont {Cho}\ \emph {et~al.}(2022)\citenamefont {Cho}, \citenamefont {Lyu}, \citenamefont {An}, \citenamefont {Han}, \citenamefont {Lo}, \citenamefont {Ng}, \citenamefont {Hu}, \citenamefont {Gao}, \citenamefont {Li}, \citenamefont {Huang}, \citenamefont {Wang}, \citenamefont {Schmalian},\ and\ \citenamefont {Lortz}}]{Cho22}%
  \BibitemOpen
  \bibfield  {author} {\bibinfo {author} {\bibfnamefont {C.-w.}\ \bibnamefont {Cho}}, \bibinfo {author} {\bibfnamefont {J.}~\bibnamefont {Lyu}}, \bibinfo {author} {\bibfnamefont {L.}~\bibnamefont {An}}, \bibinfo {author} {\bibfnamefont {T.}~\bibnamefont {Han}}, \bibinfo {author} {\bibfnamefont {K.~T.}\ \bibnamefont {Lo}}, \bibinfo {author} {\bibfnamefont {C.~Y.}\ \bibnamefont {Ng}}, \bibinfo {author} {\bibfnamefont {J.}~\bibnamefont {Hu}}, \bibinfo {author} {\bibfnamefont {Y.}~\bibnamefont {Gao}}, \bibinfo {author} {\bibfnamefont {G.}~\bibnamefont {Li}}, \bibinfo {author} {\bibfnamefont {M.}~\bibnamefont {Huang}}, \bibinfo {author} {\bibfnamefont {N.}~\bibnamefont {Wang}}, \bibinfo {author} {\bibfnamefont {J.}~\bibnamefont {Schmalian}}, \ and\ \bibinfo {author} {\bibfnamefont {R.}~\bibnamefont {Lortz}},\ }\href {\doibase 10.1103/PhysRevLett.129.087002} {\bibfield  {journal} {\bibinfo  {journal} {Phys. Rev. Lett.}\ }\textbf {\bibinfo {volume} {129}},\ \bibinfo {pages} {087002} (\bibinfo {year}
  {2022})}\BibitemShut {NoStop}%
\bibitem [{\citenamefont {Wan}\ \emph {et~al.}(2022)\citenamefont {Wan}, \citenamefont {Dreher}, \citenamefont {Mu{\~n}oz-Segovia}, \citenamefont {Harsh}, \citenamefont {Guo}, \citenamefont {Mart{\'\i}nez-Galera}, \citenamefont {Guinea}, \citenamefont {de~Juan},\ and\ \citenamefont {Ugeda}}]{Wan22}%
  \BibitemOpen
  \bibfield  {author} {\bibinfo {author} {\bibfnamefont {W.}~\bibnamefont {Wan}}, \bibinfo {author} {\bibfnamefont {P.}~\bibnamefont {Dreher}}, \bibinfo {author} {\bibfnamefont {D.}~\bibnamefont {Mu{\~n}oz-Segovia}}, \bibinfo {author} {\bibfnamefont {R.}~\bibnamefont {Harsh}}, \bibinfo {author} {\bibfnamefont {H.}~\bibnamefont {Guo}}, \bibinfo {author} {\bibfnamefont {A.~J.}\ \bibnamefont {Mart{\'\i}nez-Galera}}, \bibinfo {author} {\bibfnamefont {F.}~\bibnamefont {Guinea}}, \bibinfo {author} {\bibfnamefont {F.}~\bibnamefont {de~Juan}}, \ and\ \bibinfo {author} {\bibfnamefont {M.~M.}\ \bibnamefont {Ugeda}},\ }\href@noop {} {\bibfield  {journal} {\bibinfo  {journal} {Advanced Materials}\ }\textbf {\bibinfo {volume} {34}},\ \bibinfo {pages} {2206078} (\bibinfo {year} {2022})}\BibitemShut {NoStop}%
\bibitem [{\citenamefont {Kuzmanovi\ifmmode~\acute{c}\else \'{c}\fi{}}\ \emph {et~al.}(2022)\citenamefont {Kuzmanovi\ifmmode~\acute{c}\else \'{c}\fi{}}, \citenamefont {Dvir}, \citenamefont {LeBoeuf}, \citenamefont {Ili\ifmmode~\acute{c}\else \'{c}\fi{}}, \citenamefont {Haim}, \citenamefont {M\"ockli}, \citenamefont {Kramer}, \citenamefont {Khodas}, \citenamefont {Houzet}, \citenamefont {Meyer}, \citenamefont {Aprili}, \citenamefont {Steinberg},\ and\ \citenamefont {Quay}}]{Kuzmanovic22}%
  \BibitemOpen
  \bibfield  {author} {\bibinfo {author} {\bibfnamefont {M.}~\bibnamefont {Kuzmanovi\ifmmode~\acute{c}\else \'{c}\fi{}}}, \bibinfo {author} {\bibfnamefont {T.}~\bibnamefont {Dvir}}, \bibinfo {author} {\bibfnamefont {D.}~\bibnamefont {LeBoeuf}}, \bibinfo {author} {\bibfnamefont {S.}~\bibnamefont {Ili\ifmmode~\acute{c}\else \'{c}\fi{}}}, \bibinfo {author} {\bibfnamefont {M.}~\bibnamefont {Haim}}, \bibinfo {author} {\bibfnamefont {D.}~\bibnamefont {M\"ockli}}, \bibinfo {author} {\bibfnamefont {S.}~\bibnamefont {Kramer}}, \bibinfo {author} {\bibfnamefont {M.}~\bibnamefont {Khodas}}, \bibinfo {author} {\bibfnamefont {M.}~\bibnamefont {Houzet}}, \bibinfo {author} {\bibfnamefont {J.~S.}\ \bibnamefont {Meyer}}, \bibinfo {author} {\bibfnamefont {M.}~\bibnamefont {Aprili}}, \bibinfo {author} {\bibfnamefont {H.}~\bibnamefont {Steinberg}}, \ and\ \bibinfo {author} {\bibfnamefont {C.~H.~L.}\ \bibnamefont {Quay}},\ }\href {\doibase 10.1103/PhysRevB.106.184514} {\bibfield  {journal} {\bibinfo  {journal} {Phys. Rev. B}\
  }\textbf {\bibinfo {volume} {106}},\ \bibinfo {pages} {184514} (\bibinfo {year} {2022})}\BibitemShut {NoStop}%
\bibitem [{\citenamefont {Kezilebieke}\ \emph {et~al.}(2020)\citenamefont {Kezilebieke}, \citenamefont {Huda}, \citenamefont {Va{\v{n}}o}, \citenamefont {Aapro}, \citenamefont {Ganguli}, \citenamefont {Silveira}, \citenamefont {G{\l}odzik}, \citenamefont {Foster}, \citenamefont {Ojanen},\ and\ \citenamefont {Liljeroth}}]{Kezilebieke20}%
  \BibitemOpen
  \bibfield  {author} {\bibinfo {author} {\bibfnamefont {S.}~\bibnamefont {Kezilebieke}}, \bibinfo {author} {\bibfnamefont {M.~N.}\ \bibnamefont {Huda}}, \bibinfo {author} {\bibfnamefont {V.}~\bibnamefont {Va{\v{n}}o}}, \bibinfo {author} {\bibfnamefont {M.}~\bibnamefont {Aapro}}, \bibinfo {author} {\bibfnamefont {S.~C.}\ \bibnamefont {Ganguli}}, \bibinfo {author} {\bibfnamefont {O.~J.}\ \bibnamefont {Silveira}}, \bibinfo {author} {\bibfnamefont {S.}~\bibnamefont {G{\l}odzik}}, \bibinfo {author} {\bibfnamefont {A.~S.}\ \bibnamefont {Foster}}, \bibinfo {author} {\bibfnamefont {T.}~\bibnamefont {Ojanen}}, \ and\ \bibinfo {author} {\bibfnamefont {P.}~\bibnamefont {Liljeroth}},\ }\href@noop {} {\bibfield  {journal} {\bibinfo  {journal} {Nature}\ }\textbf {\bibinfo {volume} {588}},\ \bibinfo {pages} {424} (\bibinfo {year} {2020})}\BibitemShut {NoStop}%
\bibitem [{\citenamefont {Di~Salvo}\ \emph {et~al.}(1973)\citenamefont {Di~Salvo}, \citenamefont {Bagley}, \citenamefont {Voorhoeve},\ and\ \citenamefont {Waszczak}}]{DiSalvo73}%
  \BibitemOpen
  \bibfield  {author} {\bibinfo {author} {\bibfnamefont {F.}~\bibnamefont {Di~Salvo}}, \bibinfo {author} {\bibfnamefont {B.}~\bibnamefont {Bagley}}, \bibinfo {author} {\bibfnamefont {J.}~\bibnamefont {Voorhoeve}}, \ and\ \bibinfo {author} {\bibfnamefont {J.}~\bibnamefont {Waszczak}},\ }\href@noop {} {\bibfield  {journal} {\bibinfo  {journal} {Journal of Physics and Chemistry of Solids}\ }\textbf {\bibinfo {volume} {34}},\ \bibinfo {pages} {1357} (\bibinfo {year} {1973})}\BibitemShut {NoStop}%
\bibitem [{\citenamefont {Ribak}\ \emph {et~al.}(2020)\citenamefont {Ribak}, \citenamefont {Skiff}, \citenamefont {Mograbi}, \citenamefont {Rout}, \citenamefont {Fischer}, \citenamefont {Ruhman}, \citenamefont {Chashka}, \citenamefont {Dagan},\ and\ \citenamefont {Kanigel}}]{Ribak20}%
  \BibitemOpen
  \bibfield  {author} {\bibinfo {author} {\bibfnamefont {A.}~\bibnamefont {Ribak}}, \bibinfo {author} {\bibfnamefont {R.~M.}\ \bibnamefont {Skiff}}, \bibinfo {author} {\bibfnamefont {M.}~\bibnamefont {Mograbi}}, \bibinfo {author} {\bibfnamefont {P.}~\bibnamefont {Rout}}, \bibinfo {author} {\bibfnamefont {M.}~\bibnamefont {Fischer}}, \bibinfo {author} {\bibfnamefont {J.}~\bibnamefont {Ruhman}}, \bibinfo {author} {\bibfnamefont {K.}~\bibnamefont {Chashka}}, \bibinfo {author} {\bibfnamefont {Y.}~\bibnamefont {Dagan}}, \ and\ \bibinfo {author} {\bibfnamefont {A.}~\bibnamefont {Kanigel}},\ }\href@noop {} {\bibfield  {journal} {\bibinfo  {journal} {Science advances}\ }\textbf {\bibinfo {volume} {6}},\ \bibinfo {pages} {eaax9480} (\bibinfo {year} {2020})}\BibitemShut {NoStop}%
\bibitem [{\citenamefont {Persky}\ \emph {et~al.}(2022)\citenamefont {Persky}, \citenamefont {Bj{\o}rlig}, \citenamefont {Feldman}, \citenamefont {Almoalem}, \citenamefont {Altman}, \citenamefont {Berg}, \citenamefont {Kimchi}, \citenamefont {Ruhman}, \citenamefont {Kanigel},\ and\ \citenamefont {Kalisky}}]{Persky22}%
  \BibitemOpen
  \bibfield  {author} {\bibinfo {author} {\bibfnamefont {E.}~\bibnamefont {Persky}}, \bibinfo {author} {\bibfnamefont {A.~V.}\ \bibnamefont {Bj{\o}rlig}}, \bibinfo {author} {\bibfnamefont {I.}~\bibnamefont {Feldman}}, \bibinfo {author} {\bibfnamefont {A.}~\bibnamefont {Almoalem}}, \bibinfo {author} {\bibfnamefont {E.}~\bibnamefont {Altman}}, \bibinfo {author} {\bibfnamefont {E.}~\bibnamefont {Berg}}, \bibinfo {author} {\bibfnamefont {I.}~\bibnamefont {Kimchi}}, \bibinfo {author} {\bibfnamefont {J.}~\bibnamefont {Ruhman}}, \bibinfo {author} {\bibfnamefont {A.}~\bibnamefont {Kanigel}}, \ and\ \bibinfo {author} {\bibfnamefont {B.}~\bibnamefont {Kalisky}},\ }\href@noop {} {\bibfield  {journal} {\bibinfo  {journal} {Nature}\ }\textbf {\bibinfo {volume} {607}},\ \bibinfo {pages} {692} (\bibinfo {year} {2022})}\BibitemShut {NoStop}%
\bibitem [{\citenamefont {Nayak}\ \emph {et~al.}(2021)\citenamefont {Nayak}, \citenamefont {Steinbok}, \citenamefont {Roet}, \citenamefont {Koo}, \citenamefont {Margalit}, \citenamefont {Feldman}, \citenamefont {Almoalem}, \citenamefont {Kanigel}, \citenamefont {Fiete}, \citenamefont {Yan} \emph {et~al.}}]{Nayak21}%
  \BibitemOpen
  \bibfield  {author} {\bibinfo {author} {\bibfnamefont {A.~K.}\ \bibnamefont {Nayak}}, \bibinfo {author} {\bibfnamefont {A.}~\bibnamefont {Steinbok}}, \bibinfo {author} {\bibfnamefont {Y.}~\bibnamefont {Roet}}, \bibinfo {author} {\bibfnamefont {J.}~\bibnamefont {Koo}}, \bibinfo {author} {\bibfnamefont {G.}~\bibnamefont {Margalit}}, \bibinfo {author} {\bibfnamefont {I.}~\bibnamefont {Feldman}}, \bibinfo {author} {\bibfnamefont {A.}~\bibnamefont {Almoalem}}, \bibinfo {author} {\bibfnamefont {A.}~\bibnamefont {Kanigel}}, \bibinfo {author} {\bibfnamefont {G.~A.}\ \bibnamefont {Fiete}}, \bibinfo {author} {\bibfnamefont {B.}~\bibnamefont {Yan}},  \emph {et~al.},\ }\href@noop {} {\bibfield  {journal} {\bibinfo  {journal} {Nature physics}\ }\textbf {\bibinfo {volume} {17}},\ \bibinfo {pages} {1413} (\bibinfo {year} {2021})}\BibitemShut {NoStop}%
\bibitem [{\citenamefont {Almoalem}\ \emph {et~al.}(2024)\citenamefont {Almoalem}, \citenamefont {Feldman}, \citenamefont {Mangel}, \citenamefont {Shlafman}, \citenamefont {Yaish}, \citenamefont {Fischer}, \citenamefont {Moshe}, \citenamefont {Ruhman},\ and\ \citenamefont {Kanigel}}]{Almoalem24}%
  \BibitemOpen
  \bibfield  {author} {\bibinfo {author} {\bibfnamefont {A.}~\bibnamefont {Almoalem}}, \bibinfo {author} {\bibfnamefont {I.}~\bibnamefont {Feldman}}, \bibinfo {author} {\bibfnamefont {I.}~\bibnamefont {Mangel}}, \bibinfo {author} {\bibfnamefont {M.}~\bibnamefont {Shlafman}}, \bibinfo {author} {\bibfnamefont {Y.~E.}\ \bibnamefont {Yaish}}, \bibinfo {author} {\bibfnamefont {M.~H.}\ \bibnamefont {Fischer}}, \bibinfo {author} {\bibfnamefont {M.}~\bibnamefont {Moshe}}, \bibinfo {author} {\bibfnamefont {J.}~\bibnamefont {Ruhman}}, \ and\ \bibinfo {author} {\bibfnamefont {A.}~\bibnamefont {Kanigel}},\ }\href@noop {} {\bibfield  {journal} {\bibinfo  {journal} {Nature Communications}\ }\textbf {\bibinfo {volume} {15}},\ \bibinfo {pages} {4623} (\bibinfo {year} {2024})}\BibitemShut {NoStop}%
\bibitem [{\citenamefont {Silber}\ \emph {et~al.}(2024)\citenamefont {Silber}, \citenamefont {Mathimalar}, \citenamefont {Mangel}, \citenamefont {Nayak}, \citenamefont {Green}, \citenamefont {Avraham}, \citenamefont {Beidenkopf}, \citenamefont {Feldman}, \citenamefont {Kanigel}, \citenamefont {Klein} \emph {et~al.}}]{Silber24}%
  \BibitemOpen
  \bibfield  {author} {\bibinfo {author} {\bibfnamefont {I.}~\bibnamefont {Silber}}, \bibinfo {author} {\bibfnamefont {S.}~\bibnamefont {Mathimalar}}, \bibinfo {author} {\bibfnamefont {I.}~\bibnamefont {Mangel}}, \bibinfo {author} {\bibfnamefont {A.}~\bibnamefont {Nayak}}, \bibinfo {author} {\bibfnamefont {O.}~\bibnamefont {Green}}, \bibinfo {author} {\bibfnamefont {N.}~\bibnamefont {Avraham}}, \bibinfo {author} {\bibfnamefont {H.}~\bibnamefont {Beidenkopf}}, \bibinfo {author} {\bibfnamefont {I.}~\bibnamefont {Feldman}}, \bibinfo {author} {\bibfnamefont {A.}~\bibnamefont {Kanigel}}, \bibinfo {author} {\bibfnamefont {A.}~\bibnamefont {Klein}},  \emph {et~al.},\ }\href@noop {} {\bibfield  {journal} {\bibinfo  {journal} {Nature Communications}\ }\textbf {\bibinfo {volume} {15}},\ \bibinfo {pages} {824} (\bibinfo {year} {2024})}\BibitemShut {NoStop}%
\bibitem [{\citenamefont {Wan}\ \emph {et~al.}(2024)\citenamefont {Wan}, \citenamefont {Qiu}, \citenamefont {Ren}, \citenamefont {Qian}, \citenamefont {Li}, \citenamefont {Xu}, \citenamefont {Zhou}, \citenamefont {Zhou}, \citenamefont {Zhou}, \citenamefont {Wang} \emph {et~al.}}]{Wan23}%
  \BibitemOpen
  \bibfield  {author} {\bibinfo {author} {\bibfnamefont {Z.}~\bibnamefont {Wan}}, \bibinfo {author} {\bibfnamefont {G.}~\bibnamefont {Qiu}}, \bibinfo {author} {\bibfnamefont {H.}~\bibnamefont {Ren}}, \bibinfo {author} {\bibfnamefont {Q.}~\bibnamefont {Qian}}, \bibinfo {author} {\bibfnamefont {Y.}~\bibnamefont {Li}}, \bibinfo {author} {\bibfnamefont {D.}~\bibnamefont {Xu}}, \bibinfo {author} {\bibfnamefont {J.}~\bibnamefont {Zhou}}, \bibinfo {author} {\bibfnamefont {J.}~\bibnamefont {Zhou}}, \bibinfo {author} {\bibfnamefont {B.}~\bibnamefont {Zhou}}, \bibinfo {author} {\bibfnamefont {L.}~\bibnamefont {Wang}},  \emph {et~al.},\ }\href@noop {} {\bibfield  {journal} {\bibinfo  {journal} {Nature}\ }\textbf {\bibinfo {volume} {632}},\ \bibinfo {pages} {69} (\bibinfo {year} {2024})}\BibitemShut {NoStop}%
\bibitem [{\citenamefont {Xi}\ \emph {et~al.}(2016)\citenamefont {Xi}, \citenamefont {Wang}, \citenamefont {Zhao}, \citenamefont {Park}, \citenamefont {Law}, \citenamefont {Berger}, \citenamefont {Forr{\'o}}, \citenamefont {Shan},\ and\ \citenamefont {Mak}}]{Xi16}%
  \BibitemOpen
  \bibfield  {author} {\bibinfo {author} {\bibfnamefont {X.}~\bibnamefont {Xi}}, \bibinfo {author} {\bibfnamefont {Z.}~\bibnamefont {Wang}}, \bibinfo {author} {\bibfnamefont {W.}~\bibnamefont {Zhao}}, \bibinfo {author} {\bibfnamefont {J.-H.}\ \bibnamefont {Park}}, \bibinfo {author} {\bibfnamefont {K.~T.}\ \bibnamefont {Law}}, \bibinfo {author} {\bibfnamefont {H.}~\bibnamefont {Berger}}, \bibinfo {author} {\bibfnamefont {L.}~\bibnamefont {Forr{\'o}}}, \bibinfo {author} {\bibfnamefont {J.}~\bibnamefont {Shan}}, \ and\ \bibinfo {author} {\bibfnamefont {K.~F.}\ \bibnamefont {Mak}},\ }\href@noop {} {\bibfield  {journal} {\bibinfo  {journal} {Nature Physics}\ }\textbf {\bibinfo {volume} {12}},\ \bibinfo {pages} {139} (\bibinfo {year} {2016})}\BibitemShut {NoStop}%
\bibitem [{\citenamefont {Sohn}\ \emph {et~al.}(2018)\citenamefont {Sohn}, \citenamefont {Xi}, \citenamefont {He}, \citenamefont {Jiang}, \citenamefont {Wang}, \citenamefont {Kang}, \citenamefont {Park}, \citenamefont {Berger}, \citenamefont {Forr{\'o}}, \citenamefont {Law} \emph {et~al.}}]{Sohn18}%
  \BibitemOpen
  \bibfield  {author} {\bibinfo {author} {\bibfnamefont {E.}~\bibnamefont {Sohn}}, \bibinfo {author} {\bibfnamefont {X.}~\bibnamefont {Xi}}, \bibinfo {author} {\bibfnamefont {W.-Y.}\ \bibnamefont {He}}, \bibinfo {author} {\bibfnamefont {S.}~\bibnamefont {Jiang}}, \bibinfo {author} {\bibfnamefont {Z.}~\bibnamefont {Wang}}, \bibinfo {author} {\bibfnamefont {K.}~\bibnamefont {Kang}}, \bibinfo {author} {\bibfnamefont {J.-H.}\ \bibnamefont {Park}}, \bibinfo {author} {\bibfnamefont {H.}~\bibnamefont {Berger}}, \bibinfo {author} {\bibfnamefont {L.}~\bibnamefont {Forr{\'o}}}, \bibinfo {author} {\bibfnamefont {K.~T.}\ \bibnamefont {Law}},  \emph {et~al.},\ }\href@noop {} {\bibfield  {journal} {\bibinfo  {journal} {Nature materials}\ }\textbf {\bibinfo {volume} {17}},\ \bibinfo {pages} {504} (\bibinfo {year} {2018})}\BibitemShut {NoStop}%
\bibitem [{\citenamefont {De~la Barrera}\ \emph {et~al.}(2018)\citenamefont {De~la Barrera}, \citenamefont {Sinko}, \citenamefont {Gopalan}, \citenamefont {Sivadas}, \citenamefont {Seyler}, \citenamefont {Watanabe}, \citenamefont {Taniguchi}, \citenamefont {Tsen}, \citenamefont {Xu}, \citenamefont {Xiao} \emph {et~al.}}]{Barrera18}%
  \BibitemOpen
  \bibfield  {author} {\bibinfo {author} {\bibfnamefont {S.~C.}\ \bibnamefont {De~la Barrera}}, \bibinfo {author} {\bibfnamefont {M.~R.}\ \bibnamefont {Sinko}}, \bibinfo {author} {\bibfnamefont {D.~P.}\ \bibnamefont {Gopalan}}, \bibinfo {author} {\bibfnamefont {N.}~\bibnamefont {Sivadas}}, \bibinfo {author} {\bibfnamefont {K.~L.}\ \bibnamefont {Seyler}}, \bibinfo {author} {\bibfnamefont {K.}~\bibnamefont {Watanabe}}, \bibinfo {author} {\bibfnamefont {T.}~\bibnamefont {Taniguchi}}, \bibinfo {author} {\bibfnamefont {A.~W.}\ \bibnamefont {Tsen}}, \bibinfo {author} {\bibfnamefont {X.}~\bibnamefont {Xu}}, \bibinfo {author} {\bibfnamefont {D.}~\bibnamefont {Xiao}},  \emph {et~al.},\ }\href@noop {} {\bibfield  {journal} {\bibinfo  {journal} {Nature communications}\ }\textbf {\bibinfo {volume} {9}},\ \bibinfo {pages} {1427} (\bibinfo {year} {2018})}\BibitemShut {NoStop}%
\bibitem [{\citenamefont {Wickramaratne}\ \emph {et~al.}(2020)\citenamefont {Wickramaratne}, \citenamefont {Khmelevskyi}, \citenamefont {Agterberg},\ and\ \citenamefont {Mazin}}]{Wickramaratne20}%
  \BibitemOpen
  \bibfield  {author} {\bibinfo {author} {\bibfnamefont {D.}~\bibnamefont {Wickramaratne}}, \bibinfo {author} {\bibfnamefont {S.}~\bibnamefont {Khmelevskyi}}, \bibinfo {author} {\bibfnamefont {D.~F.}\ \bibnamefont {Agterberg}}, \ and\ \bibinfo {author} {\bibfnamefont {I.~I.}\ \bibnamefont {Mazin}},\ }\href {\doibase 10.1103/PhysRevX.10.041003} {\bibfield  {journal} {\bibinfo  {journal} {Phys. Rev. X}\ }\textbf {\bibinfo {volume} {10}},\ \bibinfo {pages} {041003} (\bibinfo {year} {2020})}\BibitemShut {NoStop}%
\bibitem [{\citenamefont {He}\ \emph {et~al.}(2018)\citenamefont {He}, \citenamefont {Zhou}, \citenamefont {He}, \citenamefont {Yuan}, \citenamefont {Zhang},\ and\ \citenamefont {Law}}]{He18}%
  \BibitemOpen
  \bibfield  {author} {\bibinfo {author} {\bibfnamefont {W.-Y.}\ \bibnamefont {He}}, \bibinfo {author} {\bibfnamefont {B.~T.}\ \bibnamefont {Zhou}}, \bibinfo {author} {\bibfnamefont {J.~J.}\ \bibnamefont {He}}, \bibinfo {author} {\bibfnamefont {N.~F.}\ \bibnamefont {Yuan}}, \bibinfo {author} {\bibfnamefont {T.}~\bibnamefont {Zhang}}, \ and\ \bibinfo {author} {\bibfnamefont {K.~T.}\ \bibnamefont {Law}},\ }\href@noop {} {\bibfield  {journal} {\bibinfo  {journal} {Communications Physics}\ }\textbf {\bibinfo {volume} {1}},\ \bibinfo {pages} {40} (\bibinfo {year} {2018})}\BibitemShut {NoStop}%
\bibitem [{\citenamefont {M\"ockli}\ and\ \citenamefont {Khodas}(2018)}]{Mockli18}%
  \BibitemOpen
  \bibfield  {author} {\bibinfo {author} {\bibfnamefont {D.}~\bibnamefont {M\"ockli}}\ and\ \bibinfo {author} {\bibfnamefont {M.}~\bibnamefont {Khodas}},\ }\href {\doibase 10.1103/PhysRevB.98.144518} {\bibfield  {journal} {\bibinfo  {journal} {Phys. Rev. B}\ }\textbf {\bibinfo {volume} {98}},\ \bibinfo {pages} {144518} (\bibinfo {year} {2018})}\BibitemShut {NoStop}%
\bibitem [{\citenamefont {Shaffer}\ \emph {et~al.}(2020)\citenamefont {Shaffer}, \citenamefont {Kang}, \citenamefont {Burnell},\ and\ \citenamefont {Fernandes}}]{Shaffer20}%
  \BibitemOpen
  \bibfield  {author} {\bibinfo {author} {\bibfnamefont {D.}~\bibnamefont {Shaffer}}, \bibinfo {author} {\bibfnamefont {J.}~\bibnamefont {Kang}}, \bibinfo {author} {\bibfnamefont {F.~J.}\ \bibnamefont {Burnell}}, \ and\ \bibinfo {author} {\bibfnamefont {R.~M.}\ \bibnamefont {Fernandes}},\ }\href {\doibase 10.1103/PhysRevB.101.224503} {\bibfield  {journal} {\bibinfo  {journal} {Phys. Rev. B}\ }\textbf {\bibinfo {volume} {101}},\ \bibinfo {pages} {224503} (\bibinfo {year} {2020})}\BibitemShut {NoStop}%
\bibitem [{\citenamefont {Margalit}\ \emph {et~al.}(2021)\citenamefont {Margalit}, \citenamefont {Berg},\ and\ \citenamefont {Oreg}}]{Margalit21}%
  \BibitemOpen
  \bibfield  {author} {\bibinfo {author} {\bibfnamefont {G.}~\bibnamefont {Margalit}}, \bibinfo {author} {\bibfnamefont {E.}~\bibnamefont {Berg}}, \ and\ \bibinfo {author} {\bibfnamefont {Y.}~\bibnamefont {Oreg}},\ }\href@noop {} {\bibfield  {journal} {\bibinfo  {journal} {Annals of Physics}\ }\textbf {\bibinfo {volume} {435}},\ \bibinfo {pages} {168561} (\bibinfo {year} {2021})}\BibitemShut {NoStop}%
\bibitem [{\citenamefont {Dentelski}\ \emph {et~al.}(2021)\citenamefont {Dentelski}, \citenamefont {Day-Roberts}, \citenamefont {Birol}, \citenamefont {Fernandes},\ and\ \citenamefont {Ruhman}}]{Dentelski21}%
  \BibitemOpen
  \bibfield  {author} {\bibinfo {author} {\bibfnamefont {D.}~\bibnamefont {Dentelski}}, \bibinfo {author} {\bibfnamefont {E.}~\bibnamefont {Day-Roberts}}, \bibinfo {author} {\bibfnamefont {T.}~\bibnamefont {Birol}}, \bibinfo {author} {\bibfnamefont {R.~M.}\ \bibnamefont {Fernandes}}, \ and\ \bibinfo {author} {\bibfnamefont {J.}~\bibnamefont {Ruhman}},\ }\href {\doibase 10.1103/PhysRevB.103.224522} {\bibfield  {journal} {\bibinfo  {journal} {Phys. Rev. B}\ }\textbf {\bibinfo {volume} {103}},\ \bibinfo {pages} {224522} (\bibinfo {year} {2021})}\BibitemShut {NoStop}%
\bibitem [{\citenamefont {H{\"o}rhold}\ \emph {et~al.}(2023)\citenamefont {H{\"o}rhold}, \citenamefont {Graf}, \citenamefont {Marganska},\ and\ \citenamefont {Grifoni}}]{Horhold23}%
  \BibitemOpen
  \bibfield  {author} {\bibinfo {author} {\bibfnamefont {S.}~\bibnamefont {H{\"o}rhold}}, \bibinfo {author} {\bibfnamefont {J.}~\bibnamefont {Graf}}, \bibinfo {author} {\bibfnamefont {M.}~\bibnamefont {Marganska}}, \ and\ \bibinfo {author} {\bibfnamefont {M.}~\bibnamefont {Grifoni}},\ }\href@noop {} {\bibfield  {journal} {\bibinfo  {journal} {2D Materials}\ }\textbf {\bibinfo {volume} {10}},\ \bibinfo {pages} {025008} (\bibinfo {year} {2023})}\BibitemShut {NoStop}%
\bibitem [{\citenamefont {Liu}\ \emph {et~al.}(2024)\citenamefont {Liu}, \citenamefont {Chatterjee}, \citenamefont {Scaffidi}, \citenamefont {Berg},\ and\ \citenamefont {Altman}}]{Liu24Interlayer}%
  \BibitemOpen
  \bibfield  {author} {\bibinfo {author} {\bibfnamefont {C.}~\bibnamefont {Liu}}, \bibinfo {author} {\bibfnamefont {S.}~\bibnamefont {Chatterjee}}, \bibinfo {author} {\bibfnamefont {T.}~\bibnamefont {Scaffidi}}, \bibinfo {author} {\bibfnamefont {E.}~\bibnamefont {Berg}}, \ and\ \bibinfo {author} {\bibfnamefont {E.}~\bibnamefont {Altman}},\ }\href {\doibase 10.1103/PhysRevB.110.024502} {\bibfield  {journal} {\bibinfo  {journal} {Phys. Rev. B}\ }\textbf {\bibinfo {volume} {110}},\ \bibinfo {pages} {024502} (\bibinfo {year} {2024})}\BibitemShut {NoStop}%
\bibitem [{\citenamefont {Hsu}\ \emph {et~al.}(2017)\citenamefont {Hsu}, \citenamefont {Vaezi}, \citenamefont {Fischer},\ and\ \citenamefont {Kim}}]{Hsu17}%
  \BibitemOpen
  \bibfield  {author} {\bibinfo {author} {\bibfnamefont {Y.-T.}\ \bibnamefont {Hsu}}, \bibinfo {author} {\bibfnamefont {A.}~\bibnamefont {Vaezi}}, \bibinfo {author} {\bibfnamefont {M.~H.}\ \bibnamefont {Fischer}}, \ and\ \bibinfo {author} {\bibfnamefont {E.-A.}\ \bibnamefont {Kim}},\ }\href@noop {} {\bibfield  {journal} {\bibinfo  {journal} {Nature communications}\ }\textbf {\bibinfo {volume} {8}},\ \bibinfo {pages} {1} (\bibinfo {year} {2017})}\BibitemShut {NoStop}%
\bibitem [{\citenamefont {Chen}\ \emph {et~al.}(2019)\citenamefont {Chen}, \citenamefont {Zhu}, \citenamefont {Zhou},\ and\ \citenamefont {An}}]{Chen19}%
  \BibitemOpen
  \bibfield  {author} {\bibinfo {author} {\bibfnamefont {W.}~\bibnamefont {Chen}}, \bibinfo {author} {\bibfnamefont {Q.}~\bibnamefont {Zhu}}, \bibinfo {author} {\bibfnamefont {Y.}~\bibnamefont {Zhou}}, \ and\ \bibinfo {author} {\bibfnamefont {J.}~\bibnamefont {An}},\ }\href {\doibase 10.1103/PhysRevB.100.054503} {\bibfield  {journal} {\bibinfo  {journal} {Phys. Rev. B}\ }\textbf {\bibinfo {volume} {100}},\ \bibinfo {pages} {054503} (\bibinfo {year} {2019})}\BibitemShut {NoStop}%
\bibitem [{\citenamefont {Lane}\ and\ \citenamefont {Zhu}(2022)}]{Lane22}%
  \BibitemOpen
  \bibfield  {author} {\bibinfo {author} {\bibfnamefont {C.}~\bibnamefont {Lane}}\ and\ \bibinfo {author} {\bibfnamefont {J.-X.}\ \bibnamefont {Zhu}},\ }\href {\doibase 10.1103/PhysRevMaterials.6.094001} {\bibfield  {journal} {\bibinfo  {journal} {Phys. Rev. Mater.}\ }\textbf {\bibinfo {volume} {6}},\ \bibinfo {pages} {094001} (\bibinfo {year} {2022})}\BibitemShut {NoStop}%
\bibitem [{\citenamefont {Zheng}\ \emph {et~al.}(2018)\citenamefont {Zheng}, \citenamefont {Zhou}, \citenamefont {Liu},\ and\ \citenamefont {Feng}}]{Zheng18}%
  \BibitemOpen
  \bibfield  {author} {\bibinfo {author} {\bibfnamefont {F.}~\bibnamefont {Zheng}}, \bibinfo {author} {\bibfnamefont {Z.}~\bibnamefont {Zhou}}, \bibinfo {author} {\bibfnamefont {X.}~\bibnamefont {Liu}}, \ and\ \bibinfo {author} {\bibfnamefont {J.}~\bibnamefont {Feng}},\ }\href {\doibase 10.1103/PhysRevB.97.081101} {\bibfield  {journal} {\bibinfo  {journal} {Phys. Rev. B}\ }\textbf {\bibinfo {volume} {97}},\ \bibinfo {pages} {081101} (\bibinfo {year} {2018})}\BibitemShut {NoStop}%
\bibitem [{\citenamefont {Das}\ and\ \citenamefont {Mazin}(2021)}]{Das21}%
  \BibitemOpen
  \bibfield  {author} {\bibinfo {author} {\bibfnamefont {S.}~\bibnamefont {Das}}\ and\ \bibinfo {author} {\bibfnamefont {I.~I.}\ \bibnamefont {Mazin}},\ }\href@noop {} {\bibfield  {journal} {\bibinfo  {journal} {Computational Materials Science}\ }\textbf {\bibinfo {volume} {200}},\ \bibinfo {pages} {110758} (\bibinfo {year} {2021})}\BibitemShut {NoStop}%
\bibitem [{\citenamefont {Costa}\ \emph {et~al.}(2022)\citenamefont {Costa}, \citenamefont {Costa},\ and\ \citenamefont {Fern\'andez-Rossier}}]{Costa22}%
  \BibitemOpen
  \bibfield  {author} {\bibinfo {author} {\bibfnamefont {A.~T.}\ \bibnamefont {Costa}}, \bibinfo {author} {\bibfnamefont {M.}~\bibnamefont {Costa}}, \ and\ \bibinfo {author} {\bibfnamefont {J.}~\bibnamefont {Fern\'andez-Rossier}},\ }\href {\doibase 10.1103/PhysRevB.105.224412} {\bibfield  {journal} {\bibinfo  {journal} {Phys. Rev. B}\ }\textbf {\bibinfo {volume} {105}},\ \bibinfo {pages} {224412} (\bibinfo {year} {2022})}\BibitemShut {NoStop}%
\bibitem [{\citenamefont {Roy}\ \emph {et~al.}(2024)\citenamefont {Roy}, \citenamefont {Kreisel}, \citenamefont {Andersen},\ and\ \citenamefont {Mukherjee}}]{Roy24}%
  \BibitemOpen
  \bibfield  {author} {\bibinfo {author} {\bibfnamefont {S.}~\bibnamefont {Roy}}, \bibinfo {author} {\bibfnamefont {A.}~\bibnamefont {Kreisel}}, \bibinfo {author} {\bibfnamefont {B.~M.}\ \bibnamefont {Andersen}}, \ and\ \bibinfo {author} {\bibfnamefont {S.}~\bibnamefont {Mukherjee}},\ }\href@noop {} {\bibfield  {journal} {\bibinfo  {journal} {2D Materials}\ }\textbf {\bibinfo {volume} {12}},\ \bibinfo {pages} {015004} (\bibinfo {year} {2024})}\BibitemShut {NoStop}%
\bibitem [{\citenamefont {Siegl}\ \emph {et~al.}(2025)\citenamefont {Siegl}, \citenamefont {Bleibaum}, \citenamefont {Wan}, \citenamefont {Kurpas}, \citenamefont {Schliemann}, \citenamefont {Ugeda}, \citenamefont {Marganska},\ and\ \citenamefont {Grifoni}}]{Siegl24}%
  \BibitemOpen
  \bibfield  {author} {\bibinfo {author} {\bibfnamefont {J.}~\bibnamefont {Siegl}}, \bibinfo {author} {\bibfnamefont {A.}~\bibnamefont {Bleibaum}}, \bibinfo {author} {\bibfnamefont {W.}~\bibnamefont {Wan}}, \bibinfo {author} {\bibfnamefont {M.}~\bibnamefont {Kurpas}}, \bibinfo {author} {\bibfnamefont {J.}~\bibnamefont {Schliemann}}, \bibinfo {author} {\bibfnamefont {M.~M.}\ \bibnamefont {Ugeda}}, \bibinfo {author} {\bibfnamefont {M.}~\bibnamefont {Marganska}}, \ and\ \bibinfo {author} {\bibfnamefont {M.}~\bibnamefont {Grifoni}},\ }\href@noop {} {\bibfield  {journal} {\bibinfo  {journal} {Nature Communications}\ }\textbf {\bibinfo {volume} {16}},\ \bibinfo {pages} {8228} (\bibinfo {year} {2025})}\BibitemShut {NoStop}%
\bibitem [{\citenamefont {Das}\ \emph {et~al.}(2023)\citenamefont {Das}, \citenamefont {Paudyal}, \citenamefont {Margine}, \citenamefont {Agterberg},\ and\ \citenamefont {Mazin}}]{Das23}%
  \BibitemOpen
  \bibfield  {author} {\bibinfo {author} {\bibfnamefont {S.}~\bibnamefont {Das}}, \bibinfo {author} {\bibfnamefont {H.}~\bibnamefont {Paudyal}}, \bibinfo {author} {\bibfnamefont {E.}~\bibnamefont {Margine}}, \bibinfo {author} {\bibfnamefont {D.}~\bibnamefont {Agterberg}}, \ and\ \bibinfo {author} {\bibfnamefont {I.}~\bibnamefont {Mazin}},\ }\href@noop {} {\bibfield  {journal} {\bibinfo  {journal} {npj Computational Materials}\ }\textbf {\bibinfo {volume} {9}},\ \bibinfo {pages} {66} (\bibinfo {year} {2023})}\BibitemShut {NoStop}%
\bibitem [{\citenamefont {Roy}\ \emph {et~al.}(2025)\citenamefont {Roy}, \citenamefont {Kreisel}, \citenamefont {Andersen},\ and\ \citenamefont {Mukherjee}}]{Roy25}%
  \BibitemOpen
  \bibfield  {author} {\bibinfo {author} {\bibfnamefont {S.}~\bibnamefont {Roy}}, \bibinfo {author} {\bibfnamefont {A.}~\bibnamefont {Kreisel}}, \bibinfo {author} {\bibfnamefont {B.~M.}\ \bibnamefont {Andersen}}, \ and\ \bibinfo {author} {\bibfnamefont {S.}~\bibnamefont {Mukherjee}},\ }\href@noop {} {} (\bibinfo {year} {2025}),\ \Eprint {http://arxiv.org/abs/arxiv:2509.03907} {arxiv:2509.03907} \BibitemShut {NoStop}%
\bibitem [{\citenamefont {Rodger}\ and\ \citenamefont {Nord{\'e}n}(1997)}]{Rodger97}%
  \BibitemOpen
  \bibfield  {author} {\bibinfo {author} {\bibfnamefont {A.}~\bibnamefont {Rodger}}\ and\ \bibinfo {author} {\bibfnamefont {B.}~\bibnamefont {Nord{\'e}n}},\ }\href@noop {} {\emph {\bibinfo {title} {Circular dichroism and linear dichroism}}},\ Vol.~\bibinfo {volume} {1}\ (\bibinfo  {publisher} {Oxford university press},\ \bibinfo {year} {1997})\BibitemShut {NoStop}%
\bibitem [{\citenamefont {Newnham}(2004)}]{Newnham04}%
  \BibitemOpen
  \bibfield  {author} {\bibinfo {author} {\bibfnamefont {R.~E.}\ \bibnamefont {Newnham}},\ }\href@noop {} {\emph {\bibinfo {title} {Properties of materials: anisotropy, symmetry, structure}}}\ (\bibinfo  {publisher} {OUP Oxford},\ \bibinfo {year} {2004})\BibitemShut {NoStop}%
\bibitem [{\citenamefont {Mirri}\ \emph {et~al.}(2016)\citenamefont {Mirri}, \citenamefont {Dusza}, \citenamefont {Bastelberger}, \citenamefont {Chinotti}, \citenamefont {Chu}, \citenamefont {Kuo}, \citenamefont {Fisher},\ and\ \citenamefont {Degiorgi}}]{Mirri16}%
  \BibitemOpen
  \bibfield  {author} {\bibinfo {author} {\bibfnamefont {C.}~\bibnamefont {Mirri}}, \bibinfo {author} {\bibfnamefont {A.}~\bibnamefont {Dusza}}, \bibinfo {author} {\bibfnamefont {S.}~\bibnamefont {Bastelberger}}, \bibinfo {author} {\bibfnamefont {M.}~\bibnamefont {Chinotti}}, \bibinfo {author} {\bibfnamefont {J.-H.}\ \bibnamefont {Chu}}, \bibinfo {author} {\bibfnamefont {H.-H.}\ \bibnamefont {Kuo}}, \bibinfo {author} {\bibfnamefont {I.~R.}\ \bibnamefont {Fisher}}, \ and\ \bibinfo {author} {\bibfnamefont {L.}~\bibnamefont {Degiorgi}},\ }\href {\doibase 10.1103/PhysRevB.93.085114} {\bibfield  {journal} {\bibinfo  {journal} {Phys. Rev. B}\ }\textbf {\bibinfo {volume} {93}},\ \bibinfo {pages} {085114} (\bibinfo {year} {2016})}\BibitemShut {NoStop}%
\bibitem [{\citenamefont {Xu}\ \emph {et~al.}(2022)\citenamefont {Xu}, \citenamefont {Ni}, \citenamefont {Liu}, \citenamefont {Ortiz}, \citenamefont {Deng}, \citenamefont {Wilson}, \citenamefont {Yan}, \citenamefont {Balents},\ and\ \citenamefont {Wu}}]{Xu22}%
  \BibitemOpen
  \bibfield  {author} {\bibinfo {author} {\bibfnamefont {Y.}~\bibnamefont {Xu}}, \bibinfo {author} {\bibfnamefont {Z.}~\bibnamefont {Ni}}, \bibinfo {author} {\bibfnamefont {Y.}~\bibnamefont {Liu}}, \bibinfo {author} {\bibfnamefont {B.~R.}\ \bibnamefont {Ortiz}}, \bibinfo {author} {\bibfnamefont {Q.}~\bibnamefont {Deng}}, \bibinfo {author} {\bibfnamefont {S.~D.}\ \bibnamefont {Wilson}}, \bibinfo {author} {\bibfnamefont {B.}~\bibnamefont {Yan}}, \bibinfo {author} {\bibfnamefont {L.}~\bibnamefont {Balents}}, \ and\ \bibinfo {author} {\bibfnamefont {L.}~\bibnamefont {Wu}},\ }\href@noop {} {\bibfield  {journal} {\bibinfo  {journal} {Nature physics}\ }\textbf {\bibinfo {volume} {18}},\ \bibinfo {pages} {1470} (\bibinfo {year} {2022})}\BibitemShut {NoStop}%
\bibitem [{\citenamefont {Argyres}(1955)}]{argyres1955theory}%
  \BibitemOpen
  \bibfield  {author} {\bibinfo {author} {\bibfnamefont {P.~N.}\ \bibnamefont {Argyres}},\ }\href {\doibase 10.1103/PhysRev.97.334} {\bibfield  {journal} {\bibinfo  {journal} {Phys. Rev.}\ }\textbf {\bibinfo {volume} {97}},\ \bibinfo {pages} {334} (\bibinfo {year} {1955})}\BibitemShut {NoStop}%
\bibitem [{\citenamefont {Xia}\ \emph {et~al.}(2006)\citenamefont {Xia}, \citenamefont {Maeno}, \citenamefont {Beyersdorf}, \citenamefont {Fejer},\ and\ \citenamefont {Kapitulnik}}]{Kapitulnik2006}%
  \BibitemOpen
  \bibfield  {author} {\bibinfo {author} {\bibfnamefont {J.}~\bibnamefont {Xia}}, \bibinfo {author} {\bibfnamefont {Y.}~\bibnamefont {Maeno}}, \bibinfo {author} {\bibfnamefont {P.~T.}\ \bibnamefont {Beyersdorf}}, \bibinfo {author} {\bibfnamefont {M.~M.}\ \bibnamefont {Fejer}}, \ and\ \bibinfo {author} {\bibfnamefont {A.}~\bibnamefont {Kapitulnik}},\ }\href {\doibase 10.1103/PhysRevLett.97.167002} {\bibfield  {journal} {\bibinfo  {journal} {Phys. Rev. Lett.}\ }\textbf {\bibinfo {volume} {97}},\ \bibinfo {pages} {167002} (\bibinfo {year} {2006})}\BibitemShut {NoStop}%
\bibitem [{\citenamefont {Kapitulnik}\ \emph {et~al.}(2009)\citenamefont {Kapitulnik}, \citenamefont {Xia}, \citenamefont {Schemm},\ and\ \citenamefont {Palevski}}]{Kapitulnik2009}%
  \BibitemOpen
  \bibfield  {author} {\bibinfo {author} {\bibfnamefont {A.}~\bibnamefont {Kapitulnik}}, \bibinfo {author} {\bibfnamefont {J.}~\bibnamefont {Xia}}, \bibinfo {author} {\bibfnamefont {E.}~\bibnamefont {Schemm}}, \ and\ \bibinfo {author} {\bibfnamefont {A.}~\bibnamefont {Palevski}},\ }\href {\doibase 10.1088/1367-2630/11/5/055060} {\bibfield  {journal} {\bibinfo  {journal} {New Journal of Physics}\ }\textbf {\bibinfo {volume} {11}},\ \bibinfo {pages} {055060} (\bibinfo {year} {2009})}\BibitemShut {NoStop}%
\bibitem [{\citenamefont {Schemm}\ \emph {et~al.}(2014)\citenamefont {Schemm}, \citenamefont {Gannon}, \citenamefont {Wishne}, \citenamefont {Halperin},\ and\ \citenamefont {Kapitulnik}}]{schemm14}%
  \BibitemOpen
  \bibfield  {author} {\bibinfo {author} {\bibfnamefont {E.}~\bibnamefont {Schemm}}, \bibinfo {author} {\bibfnamefont {W.}~\bibnamefont {Gannon}}, \bibinfo {author} {\bibfnamefont {C.}~\bibnamefont {Wishne}}, \bibinfo {author} {\bibfnamefont {W.}~\bibnamefont {Halperin}}, \ and\ \bibinfo {author} {\bibfnamefont {A.}~\bibnamefont {Kapitulnik}},\ }\href {\doibase 10.1126/science.1248552} {\bibfield  {journal} {\bibinfo  {journal} {Science}\ }\textbf {\bibinfo {volume} {345}},\ \bibinfo {pages} {190} (\bibinfo {year} {2014})}\BibitemShut {NoStop}%
\bibitem [{\citenamefont {Taylor}\ and\ \citenamefont {Kallin}(2012)}]{Taylor12}%
  \BibitemOpen
  \bibfield  {author} {\bibinfo {author} {\bibfnamefont {E.}~\bibnamefont {Taylor}}\ and\ \bibinfo {author} {\bibfnamefont {C.}~\bibnamefont {Kallin}},\ }\href {\doibase 10.1103/PhysRevLett.108.157001} {\bibfield  {journal} {\bibinfo  {journal} {Phys. Rev. Lett.}\ }\textbf {\bibinfo {volume} {108}},\ \bibinfo {pages} {157001} (\bibinfo {year} {2012})}\BibitemShut {NoStop}%
\bibitem [{\citenamefont {Wang}\ \emph {et~al.}(2014)\citenamefont {Wang}, \citenamefont {Chubukov},\ and\ \citenamefont {Nandkishore}}]{Wang14}%
  \BibitemOpen
  \bibfield  {author} {\bibinfo {author} {\bibfnamefont {Y.}~\bibnamefont {Wang}}, \bibinfo {author} {\bibfnamefont {A.}~\bibnamefont {Chubukov}}, \ and\ \bibinfo {author} {\bibfnamefont {R.}~\bibnamefont {Nandkishore}},\ }\href {\doibase 10.1103/PhysRevB.90.205130} {\bibfield  {journal} {\bibinfo  {journal} {Phys. Rev. B}\ }\textbf {\bibinfo {volume} {90}},\ \bibinfo {pages} {205130} (\bibinfo {year} {2014})}\BibitemShut {NoStop}%
\bibitem [{\citenamefont {Wang}\ \emph {et~al.}(2017)\citenamefont {Wang}, \citenamefont {Berlinsky}, \citenamefont {Zwicknagl},\ and\ \citenamefont {Kallin}}]{Wang17}%
  \BibitemOpen
  \bibfield  {author} {\bibinfo {author} {\bibfnamefont {Z.}~\bibnamefont {Wang}}, \bibinfo {author} {\bibfnamefont {J.}~\bibnamefont {Berlinsky}}, \bibinfo {author} {\bibfnamefont {G.}~\bibnamefont {Zwicknagl}}, \ and\ \bibinfo {author} {\bibfnamefont {C.}~\bibnamefont {Kallin}},\ }\href {\doibase 10.1103/PhysRevB.96.174511} {\bibfield  {journal} {\bibinfo  {journal} {Phys. Rev. B}\ }\textbf {\bibinfo {volume} {96}},\ \bibinfo {pages} {174511} (\bibinfo {year} {2017})}\BibitemShut {NoStop}%
\bibitem [{\citenamefont {Ahn}\ and\ \citenamefont {Nagaosa}(2021{\natexlab{a}})}]{Ahn21}%
  \BibitemOpen
  \bibfield  {author} {\bibinfo {author} {\bibfnamefont {J.}~\bibnamefont {Ahn}}\ and\ \bibinfo {author} {\bibfnamefont {N.}~\bibnamefont {Nagaosa}},\ }\href@noop {} {\bibfield  {journal} {\bibinfo  {journal} {Nature communications}\ }\textbf {\bibinfo {volume} {12}},\ \bibinfo {pages} {1617} (\bibinfo {year} {2021}{\natexlab{a}})}\BibitemShut {NoStop}%
\bibitem [{\citenamefont {Ahn}\ and\ \citenamefont {Nagaosa}(2021{\natexlab{b}})}]{Ahn21a}%
  \BibitemOpen
  \bibfield  {author} {\bibinfo {author} {\bibfnamefont {J.}~\bibnamefont {Ahn}}\ and\ \bibinfo {author} {\bibfnamefont {N.}~\bibnamefont {Nagaosa}},\ }\href {\doibase 10.1103/PhysRevB.104.L100501} {\bibfield  {journal} {\bibinfo  {journal} {Phys. Rev. B}\ }\textbf {\bibinfo {volume} {104}},\ \bibinfo {pages} {L100501} (\bibinfo {year} {2021}{\natexlab{b}})}\BibitemShut {NoStop}%
\bibitem [{\citenamefont {Chen}\ and\ \citenamefont {Huang}(2021)}]{Chen21}%
  \BibitemOpen
  \bibfield  {author} {\bibinfo {author} {\bibfnamefont {W.}~\bibnamefont {Chen}}\ and\ \bibinfo {author} {\bibfnamefont {W.}~\bibnamefont {Huang}},\ }\href {\doibase 10.1103/PhysRevResearch.3.L042018} {\bibfield  {journal} {\bibinfo  {journal} {Phys. Rev. Res.}\ }\textbf {\bibinfo {volume} {3}},\ \bibinfo {pages} {L042018} (\bibinfo {year} {2021})}\BibitemShut {NoStop}%
\bibitem [{\citenamefont {Xu}\ \emph {et~al.}(2019)\citenamefont {Xu}, \citenamefont {Morimoto},\ and\ \citenamefont {Moore}}]{Xu19}%
  \BibitemOpen
  \bibfield  {author} {\bibinfo {author} {\bibfnamefont {T.}~\bibnamefont {Xu}}, \bibinfo {author} {\bibfnamefont {T.}~\bibnamefont {Morimoto}}, \ and\ \bibinfo {author} {\bibfnamefont {J.~E.}\ \bibnamefont {Moore}},\ }\href {\doibase 10.1103/PhysRevB.100.220501} {\bibfield  {journal} {\bibinfo  {journal} {Phys. Rev. B}\ }\textbf {\bibinfo {volume} {100}},\ \bibinfo {pages} {220501} (\bibinfo {year} {2019})}\BibitemShut {NoStop}%
\bibitem [{\citenamefont {Tanaka}\ \emph {et~al.}(2023)\citenamefont {Tanaka}, \citenamefont {Watanabe},\ and\ \citenamefont {Yanase}}]{Tanaka23}%
  \BibitemOpen
  \bibfield  {author} {\bibinfo {author} {\bibfnamefont {H.}~\bibnamefont {Tanaka}}, \bibinfo {author} {\bibfnamefont {H.}~\bibnamefont {Watanabe}}, \ and\ \bibinfo {author} {\bibfnamefont {Y.}~\bibnamefont {Yanase}},\ }\href {\doibase 10.1103/PhysRevB.107.024513} {\bibfield  {journal} {\bibinfo  {journal} {Phys. Rev. B}\ }\textbf {\bibinfo {volume} {107}},\ \bibinfo {pages} {024513} (\bibinfo {year} {2023})}\BibitemShut {NoStop}%
\bibitem [{\citenamefont {Raj}\ \emph {et~al.}(2024)\citenamefont {Raj}, \citenamefont {Postlewaite}, \citenamefont {Chaudhary},\ and\ \citenamefont {Fiete}}]{Raj24}%
  \BibitemOpen
  \bibfield  {author} {\bibinfo {author} {\bibfnamefont {A.}~\bibnamefont {Raj}}, \bibinfo {author} {\bibfnamefont {A.}~\bibnamefont {Postlewaite}}, \bibinfo {author} {\bibfnamefont {S.}~\bibnamefont {Chaudhary}}, \ and\ \bibinfo {author} {\bibfnamefont {G.~A.}\ \bibnamefont {Fiete}},\ }\href {\doibase 10.1103/PhysRevB.109.184514} {\bibfield  {journal} {\bibinfo  {journal} {Phys. Rev. B}\ }\textbf {\bibinfo {volume} {109}},\ \bibinfo {pages} {184514} (\bibinfo {year} {2024})}\BibitemShut {NoStop}%
\bibitem [{\citenamefont {Liu}\ \emph {et~al.}(2013)\citenamefont {Liu}, \citenamefont {Shan}, \citenamefont {Yao}, \citenamefont {Yao},\ and\ \citenamefont {Xiao}}]{Liu13}%
  \BibitemOpen
  \bibfield  {author} {\bibinfo {author} {\bibfnamefont {G.-B.}\ \bibnamefont {Liu}}, \bibinfo {author} {\bibfnamefont {W.-Y.}\ \bibnamefont {Shan}}, \bibinfo {author} {\bibfnamefont {Y.}~\bibnamefont {Yao}}, \bibinfo {author} {\bibfnamefont {W.}~\bibnamefont {Yao}}, \ and\ \bibinfo {author} {\bibfnamefont {D.}~\bibnamefont {Xiao}},\ }\href {\doibase 10.1103/PhysRevB.88.085433} {\bibfield  {journal} {\bibinfo  {journal} {Phys. Rev. B}\ }\textbf {\bibinfo {volume} {88}},\ \bibinfo {pages} {085433} (\bibinfo {year} {2013})}\BibitemShut {NoStop}%
\bibitem [{\citenamefont {Schnyder}\ \emph {et~al.}(2008)\citenamefont {Schnyder}, \citenamefont {Ryu}, \citenamefont {Furusaki},\ and\ \citenamefont {Ludwig}}]{Schnyder08}%
  \BibitemOpen
  \bibfield  {author} {\bibinfo {author} {\bibfnamefont {A.~P.}\ \bibnamefont {Schnyder}}, \bibinfo {author} {\bibfnamefont {S.}~\bibnamefont {Ryu}}, \bibinfo {author} {\bibfnamefont {A.}~\bibnamefont {Furusaki}}, \ and\ \bibinfo {author} {\bibfnamefont {A.~W.~W.}\ \bibnamefont {Ludwig}},\ }\href {\doibase 10.1103/PhysRevB.78.195125} {\bibfield  {journal} {\bibinfo  {journal} {Phys. Rev. B}\ }\textbf {\bibinfo {volume} {78}},\ \bibinfo {pages} {195125} (\bibinfo {year} {2008})}\BibitemShut {NoStop}%
\bibitem [{\citenamefont {Watanabe}\ and\ \citenamefont {Watanabe}(2024)}]{watanabe_gauge-invariant_2024}%
  \BibitemOpen
  \bibfield  {author} {\bibinfo {author} {\bibfnamefont {S.}~\bibnamefont {Watanabe}}\ and\ \bibinfo {author} {\bibfnamefont {H.}~\bibnamefont {Watanabe}},\ }\href {\doibase 10.48550/arXiv.2410.18679} {{\enquote {\bibinfo {title} {A gauge-invariant formulation of optical responses in superconductors},}\ }} (\bibinfo {year} {2024}),\ \bibinfo {note} {arXiv:2410.18679 [cond-mat]}\BibitemShut {NoStop}%
\bibitem [{\citenamefont {Harrison}(2018)}]{harrison2018development}%
  \BibitemOpen
  \bibfield  {author} {\bibinfo {author} {\bibfnamefont {P.}~\bibnamefont {Harrison}},\ }\href@noop {} {\emph {\bibinfo {title} {The Development of Reflection Anisotropy Spectroscopy Instrumentation for the study of dynamic surface properties}}}\ (\bibinfo  {publisher} {The University of Liverpool (United Kingdom)},\ \bibinfo {year} {2018})\BibitemShut {NoStop}%
\bibitem [{\citenamefont {Thewalt}\ \emph {et~al.}(2018)\citenamefont {Thewalt}, \citenamefont {Hayes}, \citenamefont {Hinton}, \citenamefont {Little}, \citenamefont {Patankar}, \citenamefont {Wu}, \citenamefont {Helm}, \citenamefont {Stan}, \citenamefont {Tamura}, \citenamefont {Analytis},\ and\ \citenamefont {Orenstein}}]{PhysRevLett.121.027001}%
  \BibitemOpen
  \bibfield  {author} {\bibinfo {author} {\bibfnamefont {E.}~\bibnamefont {Thewalt}}, \bibinfo {author} {\bibfnamefont {I.~M.}\ \bibnamefont {Hayes}}, \bibinfo {author} {\bibfnamefont {J.~P.}\ \bibnamefont {Hinton}}, \bibinfo {author} {\bibfnamefont {A.}~\bibnamefont {Little}}, \bibinfo {author} {\bibfnamefont {S.}~\bibnamefont {Patankar}}, \bibinfo {author} {\bibfnamefont {L.}~\bibnamefont {Wu}}, \bibinfo {author} {\bibfnamefont {T.}~\bibnamefont {Helm}}, \bibinfo {author} {\bibfnamefont {C.~V.}\ \bibnamefont {Stan}}, \bibinfo {author} {\bibfnamefont {N.}~\bibnamefont {Tamura}}, \bibinfo {author} {\bibfnamefont {J.~G.}\ \bibnamefont {Analytis}}, \ and\ \bibinfo {author} {\bibfnamefont {J.}~\bibnamefont {Orenstein}},\ }\href {\doibase 10.1103/PhysRevLett.121.027001} {\bibfield  {journal} {\bibinfo  {journal} {Phys. Rev. Lett.}\ }\textbf {\bibinfo {volume} {121}},\ \bibinfo {pages} {027001} (\bibinfo {year} {2018})}\BibitemShut {NoStop}%
\bibitem [{\citenamefont {Tse}\ and\ \citenamefont {MacDonald}(2010)}]{PhysRevLett.105.057401}%
  \BibitemOpen
  \bibfield  {author} {\bibinfo {author} {\bibfnamefont {W.-K.}\ \bibnamefont {Tse}}\ and\ \bibinfo {author} {\bibfnamefont {A.~H.}\ \bibnamefont {MacDonald}},\ }\href {\doibase 10.1103/PhysRevLett.105.057401} {\bibfield  {journal} {\bibinfo  {journal} {Phys. Rev. Lett.}\ }\textbf {\bibinfo {volume} {105}},\ \bibinfo {pages} {057401} (\bibinfo {year} {2010})}\BibitemShut {NoStop}%
\bibitem [{\citenamefont {Dresselhaus}()}]{dresselhaus_optical_nodate}%
  \BibitemOpen
  \bibfield  {author} {\bibinfo {author} {\bibfnamefont {M.~S.}\ \bibnamefont {Dresselhaus}},\ }\href@noop {} {\emph {\bibinfo {title} {Lecture notes on Solid State Physics. \emph{Part II: Optical properties of solids.}}}},\ pp.\ \bibinfo {pages} {58--63}\BibitemShut {NoStop}%
\end{thebibliography}

%

\end{document}